\documentclass[%
	reprint,
	superscriptaddress,
	showpacs,
	amsmath,amssymb,
	aps,
	pre,
	floatfix
	]{revtex4-1}

\usepackage{graphicx}
\usepackage[colorlinks]{hyperref}

\newcommand{\figref}[1]{Fig.~\ref{#1}}
\newcommand{\eqnref}[1]{eqn.~\eqref{#1}}

\newcommand{\eg}{e.g.~}
\newcommand{\ie}{i.e.~}

\newcommand{\gi}{\ensuremath{\Gamma_\mathrm{i}}}
\newcommand{\gs}{\ensuremath{\Gamma_\mathrm{s}}}
\newcommand{\e}[1]{\ensuremath{\times 10^{#1}}}

\newcommand{\x}{\ensuremath{\mathbf{x}}}
\newcommand{\deri}[2]{\frac{d #1}{d #2}}

\newcommand{\kl}{\left( \,}
\newcommand{\kr}{\, \right)}
\newcommand{\gv}[1]{\ensuremath{\mbox{\boldmath$ #1 $}}} 
\newcommand{\abs}[1]{\lVert #1 \rVert} 

\newcommand{\vek}[1]{\ensuremath{\mathbf{#1}}}
\newcommand{\imag}{\ensuremath{\mathrm{i}}}

\usepackage[usenames]{xcolor}

\newcommand{\pdf}{pdf}

\begin{document}


\title{Order-disorder transitions in a sheared many body system}

\author{Jens C. Pfeifer}
\affiliation{Fachbereich Physik, Philipps-Universit\"at Marburg, 
	35032 Marburg, Germany}
\author{Tobias Bischoff}
\affiliation{Division of Geological and Planetary Sciences, California Institute
	of Technology, Pasadena, CA, 91125,USA} 
\author{Georg Ehlers}
\affiliation{Fachbereich Physik, Philipps-Universit\"at Marburg, 
	35032 Marburg, Germany}
\author{Bruno Eckhardt}
\affiliation{Fachbereich Physik, Philipps-Universit\"at Marburg, 
	35032 Marburg, Germany}

\date{\today}

\begin{abstract}
Motivated by experiments on sheared suspensions that show a transition between
ordered and disordered phases, we here study the long-time behavior of a sheared
and overdamped 2-d system of particles interacting by repulsive forces. As a
function of interaction strength and shear rate we find transitions between
phases with vanishing and large single-particle diffusion. In the phases with
vanishing single-particle diffusion, the system evolves towards regular
lattices, usually on very slow time scales. Different lattices can be
approached, depending on interaction strength and forcing amplitude. The
disordered state appears in parameter regions where the regular lattices are
unstable. Correlation functions between the particles reveal the formation of
shear bands. In contrast to single particle densities, the spatially resolved
two-particle correlation functions vary with time and allow to determine the
phase within a period. As in the case of the suspensions, motion in the state
with low diffusivity is essentially reversible, whereas in the state with strong
diffusion it is not.
\end{abstract}




\maketitle

\section{Introduction}
The observation of echoes, i.e. a recovery of the initial state, for spins
precessing in a magnetic field upon reversal of the field
\cite{Hahn1950,Carr1954} and for dye in viscous fluids \cite{Homsy2008} are
among the best known experiments that seemingly defy the expected
irreversibility in the evolution of many body systems. Several other examples
have followed \cite{Niggemeier1993,Chaiken1986}, and in the context of chaotic
dynamical systems the absence of echoes and the inability to return to initial
conditions has frequently been used as a test for chaotic dynamics
\cite{Roberts1992,Casati1986,Pastawski1995,Eckhardt2003}. While in all these
cases the strength of the echoes deteriorates gradually as parameters are
varied, the experiments by Pine et al \cite{Pine2005} on sheared suspensions
added a new facet to the old problem: they find a phase-transition like change
from an essentially reversible state for one set of parameters to an
irreversible state for other parameters. Subsequent studies have demonstrated
similar transitional behavior in other hydrodynamic experiments
\cite{Guasto2010,Metzger2012,Jeanneret2014}, particle systems
\cite{Franceschini2011,*Franceschini2014,Keim2013a,*Keim2013}, and complementary
simulations \cite{Metzger2010} and simplified models
\cite{During2009,Corte2008,Corte2009,Menon2009,Keim2011}. It has also been put
into the context of more general order-disorder transitions in solid materials
\cite{Mohan2013,Regev2013,Fiocco2013,*Fiocco2014} and granular systems
\cite{Slotterback2012,Schreck2013,Royer2014}. Similar transitions were also
reported for superconducting vortices
\cite{Mangan2008,Zhang2010,Okuma2010,*Motohashi2011,*Okuma2011}.

The demonstration of the transition in several other systems suggests that it is
a general phenomenon for sheared many body systems. This is also supported
by the ability to capture much of the observed behavior and the transition in
the discrete time model described in \cite{Corte2008}. In order to explore the
origin of the phenomenon further and to also establish a connection to
reversibility studies in chaotic dynamical systems, we here consider a model for
a many-body system that contains some of the features of the suspension
experiment but allows for a much more detailed investigation of its long-time
properties.

In section \ref{sec:Model} we  present the details of the model and demonstrate
that it also shows the transition observed in \cite{Pine2005}. In section
\ref{sec:Order} we  characterize the system both before and after the disorder
transition. In section \ref{sec:Floquet} we  investigate the stability
properties of the limit cycles described in the previous sections. Finally, in
section \ref{sec:Dist} we  analyze the spatially resolved correlations and their
phase dependency. We conclude with a few general remarks and observations in
section \ref{sec:sum}.

\section{The model and its parameters}\label{sec:Model}

For our model we do not aim to calculate the detailed hydrodynamic interaction
or exact forces between charged colloids, as done in the simulations of
\cite{Pine2005}. Rather, we pick an interaction force that captures key features
of the fluid-particle system. This is similar in spirit to the model in
\cite{Corte2008}, but differs in that the model is a continuous one and not an
event-driven mapping. We assume that the motion of the many body system is
overdamped, so that there is no motion if the external forcing ceases. The
forces on the particles are hence balanced by viscous friction, and the
equations of motion become
\begin{equation}\label{eq:ch1eq1}
	\mu \deri{\x_i}{t} = \vek{F}_i.
\end{equation}
Here, $\mu$ is a friction coefficient, $\vek x_i$ the position of particle $i$,
and $\vek F_i$ are the forces acting on the particle. We take the interaction
between the particles as repulsive, thereby mimicking the repulsion due to
liquid pressure when two particles come close. The particles are confined to a
plane, and the domain is taken to be rectangular, and not curved as between the
cylinders in the experiments of \citet{Pine2005}. The domain is periodically
continued in both directions. The main control parameter will be the amplitude
of the forcing. A second parameter controls the strength of the interaction
between the particles; it is also related to the density of the system, as will
be discussed below.

\subsection{Forces}
The repulsion between the particles is modeled as a potential with power-law
decay, $V(r_{ij})\propto r_{ij}^{-\alpha}$, with $r_{ij}$ the distance between
particles $i$ and $j$. We here take  $\alpha=2$ as a compromise between strong
repulsion for short distances (i.e. larger $\alpha$) and a numerically
controllable range of the interaction (where hydrodynamics would suggest a decay
as slow as $\alpha=1$). With this repulsive interaction, the system is
reminiscent of Wigner crystals and clusters of charged particles in plasmas
\cite{Melzer1996,*Melzer1996a,*Schweigert1998}. Particle positions are denoted
${\bf x}_i=(x_i,y_i)$, with $x$ along the direction of shear and $y$ normal to
it.  The total force acting on the \mbox{$i$-th} particle is the sum of the mutual
interactions with all other particles in the system and a periodic shear force
in the $x$-direction whose amplitude increases linearly with $y_i$,
\begin{equation}\label{eq:ch1eq3}
\vek{F}_i = 
- \nabla \kl \sum \limits_{j \neq i}  \frac{A}{\abs{\x_i-\x_j}^2} \kr 
+ Sy_i\cos(\omega t)\vek{e}_x.
\end{equation}
The parameter $A$ determines the strength of the inter particle potential,
$Sy_i$ is the amplitude of the shear force and $\omega$ is its frequency.
Substituted into equation (\ref{eq:ch1eq1}), we obtain the full expression for
the time evolution of the position 
of the $i$-th particle,
\begin{equation}\label{eq:ch1eq4}
\deri{\x_i}{t} = 
 2\frac{A}{\mu} \sum \limits_{j \neq i}  \frac{\x_i-\x_j}{\abs{\x_i-\x_j}^4}
 + \frac{S}{\mu}y_i\cos(\omega t)\vek{e}_x \,.
\end{equation}

\subsection{Parameters}
In order to expose the independent parameters of the system, we introduce a
length scale $\lambda$ and a time scale $T$, and set $\x_i = \lambda \x_i'$ and
$t=Tt'$. The time $T$ can naturally be identified with the period of the shear,
$T=2\pi/\omega\frac{}{}$. The length scale is not related to a quantity
explicitly displayed in the equations, but it enters implicitly in the many-body
system via the mean distance between particles, and is thus related to the
density. Therefore, variations in period and density are absorbed into the two
remaining parameters of the system, the dimensionless interaction strength
\begin{equation}
\gi = 2AT/\mu\lambda^4
\end{equation} 
and the dimensionless shear rate
\begin{equation}
\gs = ST/\mu.
\end{equation}
The evolution equations then become
\begin{equation}\label{eq:ch1eq5}
\deri{\x_i'}{t} = 
 \gi \sum \limits_{j \neq i}  \frac{\x_i'-\x_j'}{\abs{\x_i'-\x_j'}^4}
 + \gs y_i'\cos(2\pi t)\vek{e}_x .
\end{equation}
We will drop the primes in the subsequent sections and refer to this equation as
the evolution equation.

The instantaneous strain $\gamma(t)$ follows from the time evolution of the
distance between two particles that initially are displaced by $\Delta y$
perpendicular to the shear. The separation in $x$-direction is then given by
$\Delta x(t)=\gamma(t) \Delta y$ with
\begin{equation}
 \gamma(t) =  \frac{\Delta x(t)}{\Delta y} = 
 \gs / (2\pi) \sin(2\pi t) = \gamma_0 \sin(2\pi t)\,;
\end{equation}
$\gamma_0=\gs/(2\pi)$ is the amplitude of the affine shear. This also gives
a notion of the shear process: At the start of the period, $t=0$, the strain is
zero. Over the next half period, the system is tilted to the right,  reaching
its apex at $t=T/4$. Here, the flow comes to a halt and reverses its direction.
After half a period, strain is again zero and the process repeats to
the left. The total accumulated strain over one period then becomes 
\begin{equation}
	\gamma = \int_0^1 \left| \gs \cos(2\pi t) \right| dt = 4 \gamma_0.
\end{equation}

\subsection{Numerical implementation}

In order to solve equation (\ref{eq:ch1eq5}), we introduce Lees-Edwards boundary
conditions \cite{Lees1972} to account for the sheared images in the
$y$-direction. We use a modified minimum image convention, taking several
closest images of each particle into account, typically one or two in each
direction. The typical number of particles in the main domain is about 100, and
in a few examples we went up to 900 particles. Time integration is performed
using a 4th order Runge-Kutta-Fehlberg integrator provided by the GNU Scientific
Library (GSL). Computation of the right-hand side is parallelized using openMP.
The width (in the $x$-direction) of the box is $K$, whereas the height (in the
$y$-direction) of the box is $\sin(\pi/3)L$ with $K,L \in \mathbb{N}$ and even
and $KL = N$, as suggested by a regular triangular lattice. The simulation was
carried out with random initial conditions, and the motion was typically
followed over several thousand periods in order to obtain clear evidence for the
asymptotic state and to extract reliable statistics in the case of chaotic
states.

\subsection{Reduced representations and stroboscopic maps}\label{ch:ch1sec3}

\begin{figure}[t]
 \includegraphics[width=\linewidth]{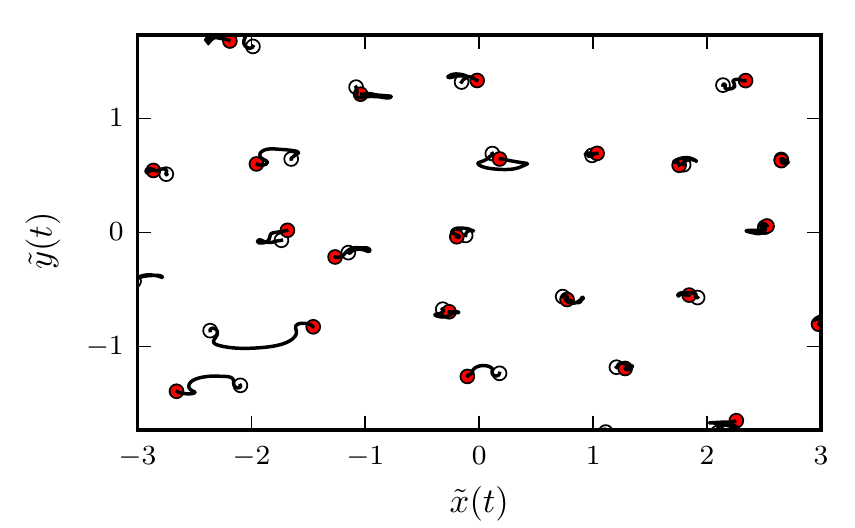}
 \caption[]{
 	(Color online)
 	Particle trajectories over one period. Initial positions are marked by open
 	circles, final positions by filled red circles. Shown are the reduced
 	positions (\ref{eq:ch1reducedx}) after subtracting the affine transformation.
 	Non-reversible motions (in particular in the lower left corner) are connected
 	with close encounters and large displacements, see also supplementary movie
 	online \cite{Note1}. The parameters are $\gi = 0.1$ and $\gs = 16.0$.
 \label{fig:ch1fig1}
 }
\end{figure}

Even for small shear, particles undergo large scale motions along the shear
axis, with an amplitude that increases with the displacement in the normal
direction. In order to remove this affine deformation, we introduce a reduced
representation where the translation by the shear is taken out, i.e. we study
\begin{equation}
\tilde x_i(t) = x_i(t) - \gamma_0 y(t_0) \sin(2\pi t) \,.
\label{eq:ch1reducedx}
\end{equation} 
Figure \ref{fig:ch1fig1} shows examples of such reduced motions: for the chosen
parameters the amplitudes of the motion around the reference positions are small
and most particles stay close to their initial positions, but as the group in
the lower left corner shows, some of them can experience large displacements as
a consequence of close encounters.

\begin{figure}
 \includegraphics[width=\linewidth]{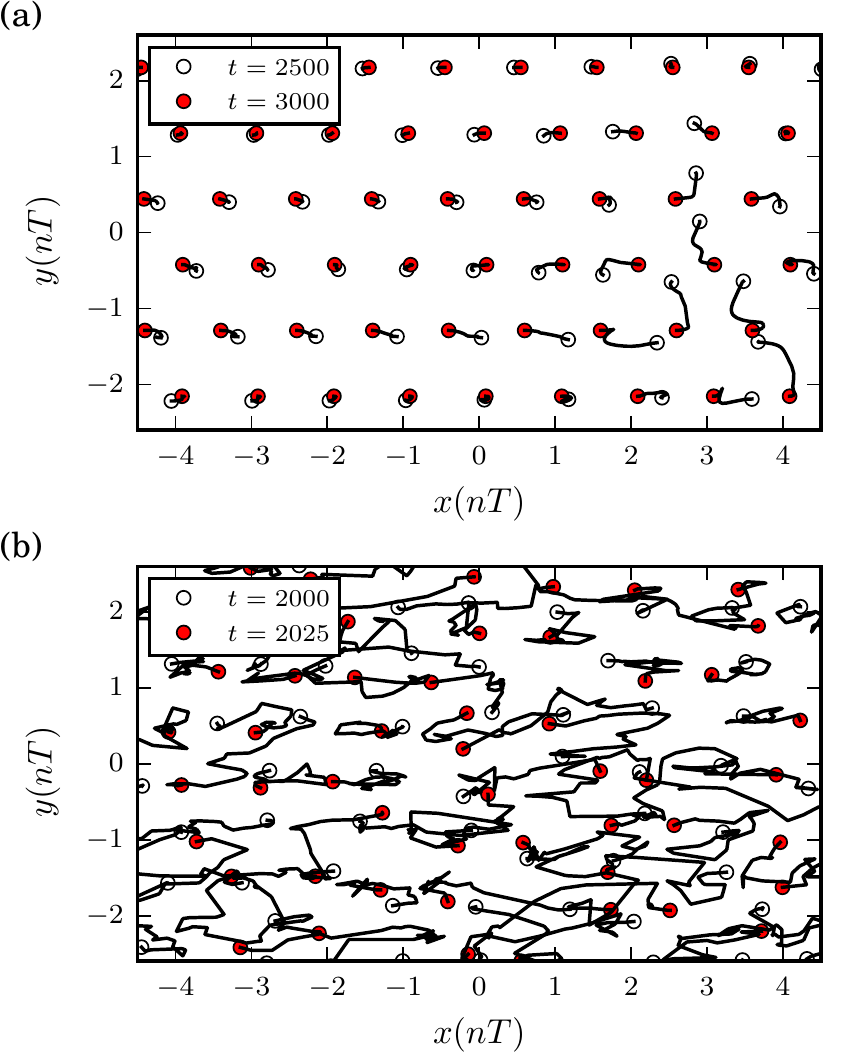}
 \caption{
 (Color online)
 Stroboscopic map of particle trajectories for low shear ($\gs=8$, top) and
 large shear ($\gs=16$, bottom). Open circles indicate initial positions,
 full red circles final positions. In the top diagram for small shear one notes
 that the particles move very little over the 500 time steps. There is a
 localized region with larger motions in the lower right corner that is
 connected with a rare local rearrangement. In the lower diagram for larger
 shear almost all particles experience large displacements over a much shorter
 interval of 25 time units. In both cases $\gi=0.1$.
 \label{fig:ch1fig2}
 }
\end{figure}

The long time behavior of the system shows up after a large number of periods
only, and is then best studied in the form of stroboscopic maps, with positions
taken at multiples of the period, ${\bf x}_i(n)={\bf x}_i(nT)$. In figure
\ref{fig:ch1fig2} and the accompanying movies \footnote{See Supplemental
	Material at URL for movies.} we show two examples, one for smaller shear rate
and one for larger shear rate. For moderate shear rates, only small
displacements are observed  and the particles evolve slowly towards regular
lattices (section \ref{ch:reg}). For larger shear, almost all particles travel
larger distances from their initial position. This corresponds to the disordered
state of \citet{Pine2005}. The random motion of the particles can be captured by
diffusion coefficients, which are anisotropic and differ in the longitudinal and
transverse directions (section \ref{sec:Diffuse}).


\section{Ordered and disordered states}\label{sec:Order}

\subsection{Ordered states}\label{ch:reg}

In the absence of external forcing, the interactions between the particles push
a random initial condition towards a force equilibrium. A regular lattice, such
as a hexagon, is an example of such an equilibrium, but it will rarely be
reached from a random initial condition since particles will be trapped in a
state with a few dislocations or in a state showing several oriented patches,
separated by grain boundaries.

\begin{figure}[t]
	\includegraphics[width=\linewidth]{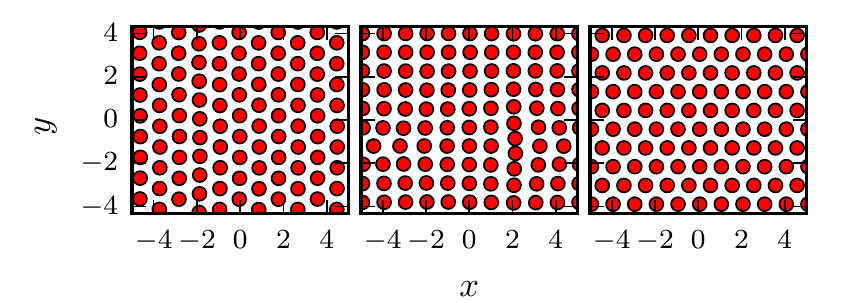}
	\caption[Asymptotic states]{
		(Color online)
		Asymptotic states at moderate shear rates, $\gs = 2$, $6$ and $8$,
		respectively. Except for minor defects they are either a hexagonal or a
		rectangular lattice configuration. $\gi = 0.1$ in all three cases.
		\label{fig:Asymptotic}}
\end{figure}

As soon as shear is added to the system, the particles will start to rearrange
and to self-organize. After sufficiently long times, states like the ones shown
in \figref{fig:Asymptotic} for $\gs = 2$, $6$, and $8$ are obtained.
For very low shear rates the system settles into the hexagonal configuration,
with an orientation relative to the shear direction that depends on the initial
conditions. Since the regular hexagonal state is a possible force-equilibrium,
the external forcing can be seen as a minor perturbation that hardly affects
particle interactions. With increasing shear rate, the lattice orientation may
change, and the lattice will align parallel or perpendicular to the external
force. At a shear rate around $6$, the asymptotic state is given by a
rectangular lattice configuration. At $\gs \approx 8$, we again observe a
hexagonal lattice, though this time it is almost always oriented parallel to the
shearing motion. Increasing the rate further the system will at some point fail
to approach a regular lattice, and a chaotic, disordered state is attained.

Since the asymptotic reversible states are also lattice configurations, the
external forcing only causes affine deformations of the system. This is in
contrast to the observations by \citet{Keim2013} who found clusters of particles
subject to non-affine deformations in the reversible regime. We only observe
non-affine deformations in the irreversible regime and in the trajectories
approaching the ordered states.

The regular states observed here are a consequence of the homogeneity of the
system, the density, and the domain size: Imperfections will most likely trap
the system in a glassy or disordered state, as in the experiments of
\citet{Pine2005}. Another important aspect is the different kind of interactions
considered in \cite{Pine2005,Corte2008}. Since there an interaction between
particles takes place upon contact only, reversibility is attained as soon as
each particle has  cleared a sufficiently large space around it. In our case,
particles interact over longer distances so that further ordering is required
and eventually ordered states are attained. Nevertheless, as \citet{Corte2008}
pointed out, very similar absorbing states exist in their systems as well, but
are not found by the simulations. Similar to our case, they become unstable as
soon as a critical strain is exceeded.

In order to quantify the ordering we investigate the orientation number of an
ensemble, defined by
\begin{equation}
	\Psi_k = \left\langle \frac{1}{k}\sum_{j=1}^k\exp\left( \imag k\Theta_{ij} \right) \right\rangle.
\end{equation}
Here, $k$ is the number of next neighbors taken into account, and $\Theta_{ij}$
is the angle of the vector between particles $i$ and $j$ towards the $x$-axis.
Of particular interest are the values of $|\Psi_6|$, which becomes unity for a
hexagonal lattice, and  of $|\Psi_4|$, which becomes unity for a rectangular
lattice.

\begin{figure}
	\includegraphics[width=\linewidth]{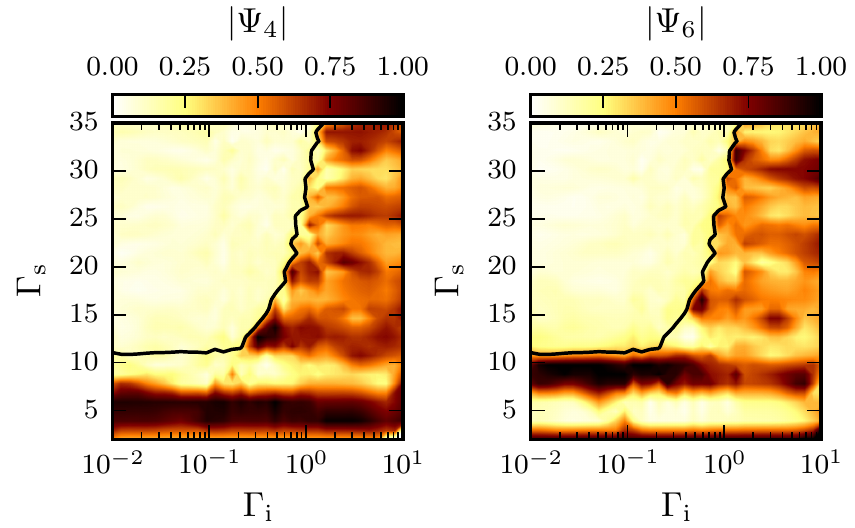}
	\caption{
		(Color online)
		Order parameters $|\Psi_4|$ and $|\Psi_6|$ for different values of $\gi$ and
		$\gs$. For a low shear rate $\gs$ we clearly observe a change from sixfold
		orientation to fourfold orientation (at $\gs \approx 3$) and back to sixfold
		orientation (at $\gs \approx 7$). In the light area in the upper left, no
		ordering of any kind is observable. Its lower boundary is located near $\gs
		\approx 11$. For larger $\gi$, the system attains lattice configurations even
		at larger shear rates. Although the system then does not reach perfect lattice
		configurations, a modulation with $\gs$ is still discernible. The black solid
		lines indicate the loss of ordering where both $|\Psi_4|$ and $|\Psi_6|$ drop
		below $0.4$.
		\label{fig:Ordering}}
\end{figure}

We show the results in \figref{fig:Ordering} for a wide range in parameter space
$\gi \times \gs$. For low shear rates and independent of the interaction
strength, the orientation number reflects the previously mentioned change of the
asymptotic states. For very low shear rates up to $\gs \lesssim 3$ and for shear
rates larger than $\gs \gtrsim 7$, a hexagonal lattice is preferred. For shear
rates in between we find $|\Psi_4| \approx 1$, and thus a rectangular lattice is
attained. With increasing shear rate and low enough interaction strength $\gi$,
order is lost beyond $\gs \gtrsim 11$. This corresponds to the observation of
chaotic states such as in \figref{fig:ch1fig2}(b). For a larger interaction
strength $\gi$, ordered states are attained even at large shear rates. However,
the computational demand increases and integration times become too short, and
hence several defects remain in the configurations and reduce the orientation
number. However, in contrast to the orderless region, we find an orientation
number which is larger than $0.4$ in this region. Additionally, we observe a
modulation of the orientation number with $\gs$, in which both lattice types
alternately dominate.

\subsection{Diffusion constants}\label{sec:Diffuse}
\begin{figure}
 \includegraphics[width=\linewidth]{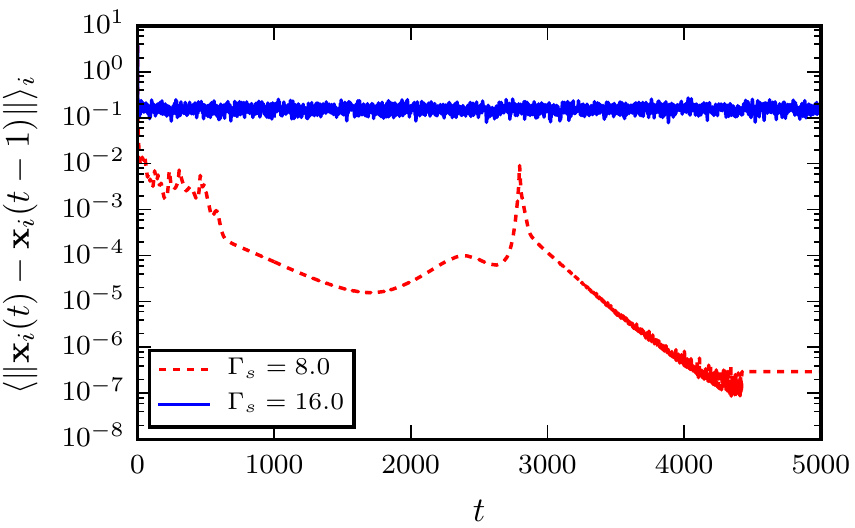}
 \caption{
 	(Color online)
	Average displacement after one period for different shear rates $\gs$. For
	low shear rates, it drops to numerical precision with intermediate bumps
	indicating reorganizations. For large shear rates, the average displacement
	approaches a finite value reflecting the random motion. $\gi = 0.1$.
 \label{fig:ch1fig4}
 }
\end{figure}

Convenient scalar quantities that can be used to distinguish the different
long-time dynamics are related to the diffusive motions of the particles. The
local and instantaneous quantity is the mean displacement of particles over one
period,
\begin{equation}
\delta s(n)= \frac{1}{N} \sum_{k=1}^N{} 
 \left\| {\bf x}_k(n) - {\bf x}_k \left(n-1\right)  \right\|
 \label{eq:local_D}
\end{equation}
It provides a convenient measure to quickly distinguish between regular and
irregular states, as shown in \figref{fig:ch1fig4}. For large shear
($\gs=16$) the displacement is large and varies little in time, indicating
a disordered, chaotic state. For low shear ($\gs=8$) it decreases,
eventually reaching machine precision: the system evolves towards a stable
state. The intermediate maxima, such as the one near $t=2800$ in
\figref{fig:ch1fig4}, indicate reorganizations to remove lattice defects. The
event is related to the rearrangement in the lower right corner in
\figref{fig:ch1fig2} (top) and clearly visible in the movie \cite{Note1}.

In the diffusive state, we can calculate diffusion constants. In view of the
asymmetry in the system in the directions along the shear and perpendicular to
it, we consider two separate diffusions constants. The diffusion constants can
be obtained either by averaging in time over the displacements over one period,
\begin{equation}\label{eq:ch2eq1}
2D_x \gamma
= \left\langle   \frac{1}{N} \sum_{k=1}^N{} \lvert x_k(n) - x_k\left(n-1\right)
 \rvert^2\right\rangle_{n>n_{\mathrm{transient}}} 
\end{equation}
or by approximating the mean square displacement of the particles by a straight
line,
\begin{equation}\label{eq:ch2eq2}
 \langle \Delta x^2(n) \rangle = 2D_x \gamma n.
\end{equation}
The factor $\gamma$ is adopted from the definitions in \cite{Pine2005}.
Expressions for the diffusion constant in the $y$-direction are obtained by
replacing $x_k$ by $y_k$. Due to the slightly different definitions, we call
\eqnref{eq:ch2eq1} short term diffusivity, and \eqnref{eq:ch2eq2} the long term
diffusivity. The averages are taken after initial transients have decayed. The
number of periods one has to wait depends on the system size and the asymptotic
behavior: While in the disordered regime, the system approaches its final state
quickly, this time may be several thousand periods when the system approaches an
ordered configuration. This shows up in $\delta s$, which can be used to
distinguish ordered from disordered states. For disordered states, the diffusion
constant is obtained by averaging over $10^3$ to $10^4$ periods. In the ordered
states, the diffusion constant calculated from averages over shorter time
intervals shows a drop towards zero.

Since the simulation of large ensembles is computationally expensive, we
concentrate on a system size of $N=100$ from here on. To verify that the results
are not affected by the system size, we occasionally investigate systems of
sizes $N=400$ and $900$ for a few parameter values. While the results for the
diffusion constants seem to be independent of the system size as shown below,
the time for transients to decay increases rapidly with system size and adds to
the numerical challenges.

\subsection{Transitions for fixed $\gi$}
\begin{figure}
	\includegraphics[width=\linewidth]{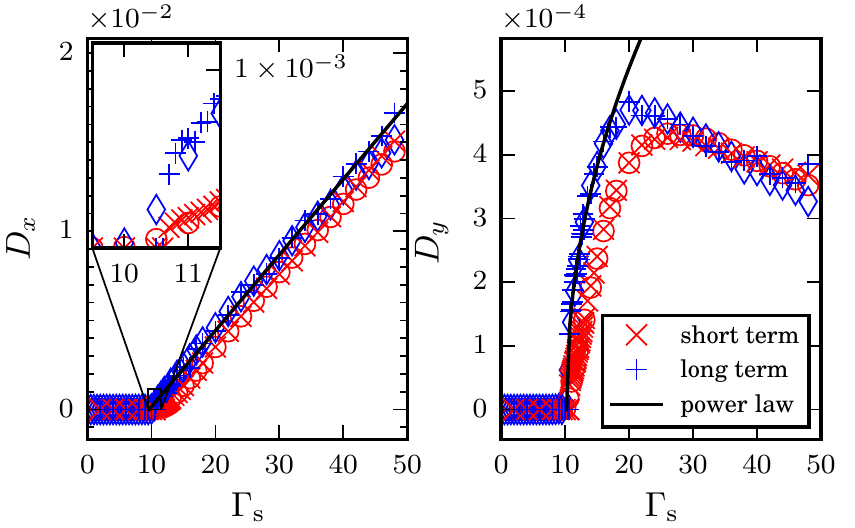}
	\caption[Diffusion constants, slice]{
		(Color online)
		Diffusivities along the $x$- and $y$-direction, left and right panel
		respectively, for $\gi = 0.1$. Additional data of a 900-particle systems is
		marked by open symbols (short term: {\color{red} $\circ$}; long term:
		{\color{blue} $\diamond$}). For low shear rates, no diffusion is observed,
		corresponding to a reversible state. At a critical point $9.5 <
		\Gamma_{2,\mathrm{c}} < 10.5$ diffusions start to grow, both parallel and
		perpendicular to the shear, corresponding to an irreversible state. The inset
		in the left panel shows a magnification of the critical region. The short term
		diffusivity is slightly smaller than the long term diffusivity. Moreover,
		diffusion parallel to the shear is strongly enhanced and shows a different
		behavior: It grows linearly with increasing shear whereas diffusion in the
		perpendicular direction saturates. Boundary effects due to the system size
		seem to be negligible up to $\gs \approx 30.0$ since data points for $100$ and
		$900$ particles coincide well.
	\label{fig:ch2fig1}}
\end{figure}

We begin the exploration of the parameter space of the system with a fixed
interaction parameter $\gi=0.1$ and different shear rates $\gs$. We
compare both short term and long term diffusivities in \figref{fig:ch2fig1}. As
anticipated from the results in the previous section, we notice two
qualitatively different states that can be distinguished by their diffusion
constants. In one case, the system approaches a state with zero diffusivity and
reversible motion. In the other case, the diffusivity does not vanish and the
motion is irreversible. For this specific set of parameters, we computed the
diffusivities both for a 100 and a 900 particle system. Up to a shear rate of
$\gs$ just above 30.0, the results are in very good agreement. Beyond this
value, the system with 100 particles gives slightly higher values. This is most
likely caused by finite size effects: At $\gs = 36$, which corresponds to a
strain amplitude $\gamma_0 = 5.73$, particles separated by $\Delta y =
\sqrt{3}/2$ in the normal direction are advected by one box length relative to
each other. Therefore, at the left and right turning points of the shearing
motion, particles experience the same neighborhood.

The transition between reversible and irreversible motion as the shear rate is
varied shows up rather clearly. For $\gi=0.1$, this transition takes place near
$\gs \approx 10.5$. Fitting a power law of the form $D = \delta\left(\gs -
\Gamma_{2,\mathrm{c}}\right)^\beta$ returns an exponent $\beta_x = 1.00$, a
critical value $\Gamma_{2,\mathrm{c}}^x = 9.6$ for the diffusivity in the shear
direction, and $\beta_y = 0.47$ and $\Gamma_{2,\mathrm{c}}^y = 10.2$ for the
diffusivity in the normal direction. The values for the critical shear rate are
lower than the point where a non-zero diffusivity is first observed. The inset
in \figref{fig:ch2fig1} shows this critical region in more detail. A closer
inspection of the diffusivities near the critical point reveals that they drop
to zero abruptly. This suggests a small region where ordered and disordered
motions coexist, so that the transition could be subcritical and hysteresis
effects might be observed. Unexpected are the differences in the exponents:
Parallel to the shear direction, the diffusivity grows linearly with the shear,
but perpendicular to it, the diffusivity varies with a square root. Overall, the
sharp transition is in qualitative agreement with the results both from
experiments and simulations, \eg \cite{Pine2005,Metzger2010,Guasto2010}.

We find that the short term diffusivity \eqnref{eq:ch2eq1} is slightly lower
than its long term counterpart. Parallel to the shear direction, the difference
approaches a constant value, whereas in the perpendicular direction, it
diminishes for larger shear rates. Our interpretation is that on short time
scales, each particle is captured in a matrix of surrounding particles. Their
presence introduces memory and correlations into the process, thereby reducing
the diffusion coefficient. This effect is most pronounced in the parameter range
close to the transition and becomes weaker with increasing shear.

\subsection{Anisotropy in diffusion}
We also notice that horizontal diffusivities are much stronger than vertical
ones. This is due to the shear acting only along the $x$-direction and was also
observed in other investigations on similar systems
\cite{Pine2005,Metzger2010}. Moreover, for larger shear rates, the vertical
diffusivity seems to approach a fixed value whereas the horizontal diffusivity
grows linearly over the investigated parameter range. This phenomenon of enhanced
diffusion is known as advection-diffusion coupling and has been studied for the
case of Brownian motion in a shear flow by \citet{Young1982}. They showed that
the diffusion constants are related to each other by
\begin{equation}\label{eq:chSheareqCollapse}
	D_x = D_{x,0} + \frac{1}{2}\gamma_0^2 D_y,
\end{equation}
where $\gamma_0$ is the strain amplitude. Thus, parallel diffusion is enhanced
by a coupling between normal diffusion and shearing. The bare longitudinal
diffusivity without this effect is denoted $D_{x,0}$. The mechanism can be
illustrated in a three step process by decoupling diffusion and advection: In
the first step, we observe diffusion perpendicular to the shear so that
particles are found in the layer above. In the next step, this layer is advected
by the shear over some distance. In the last step, particles in the upper layer
diffuse back to the original layer. Since they have been transported by the
flow, they have traveled further than by horizontal diffusion alone. Equation
\eqref{eq:chSheareqCollapse} shows that 
for an isotropic system the bare longitudinal and transversal
diffusivities coincide.

\begin{figure}
	\center
	\includegraphics[width=\linewidth]{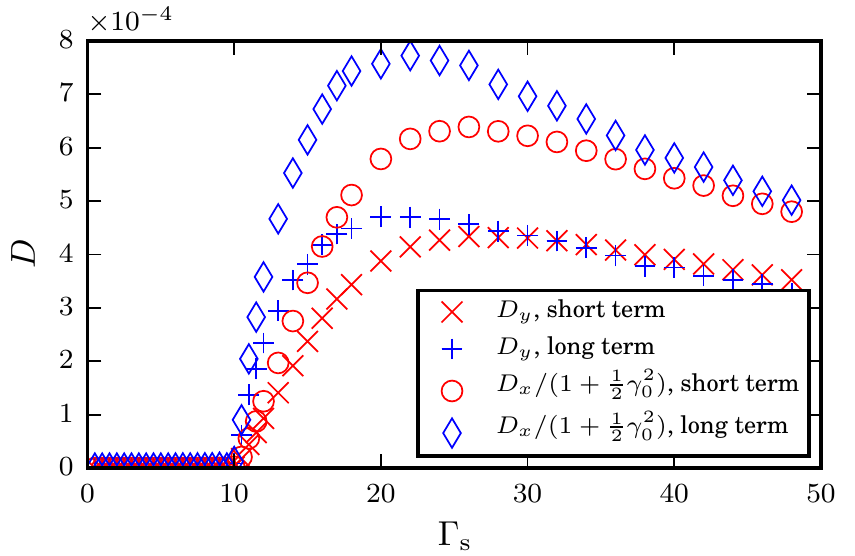}
	\caption[Diffusion constants, collapsed]{
		(Color online)
		Relation of the corrected parallel diffusivity (\eqnref{eq:chSheareqCollapse})
		to the perpendicular diffusivity. The rescaled parallel diffusivity is still
		stronger, indicating a true anisotropy. Data was taken from the 900-particle
		system.
		\label{fig:chShearfigCollapse}}
\end{figure}
 
In \figref{fig:chShearfigCollapse}, we show the corrected parallel in relation
to the perpendicular diffusivity. Data was taken with the 900-particle system.
We consider $D_x/(1+\frac{1}{2}\gamma_0^2)$, which in the limit of large
$\gamma_0$ should approach $D_y$. We observe that the bare parallel diffusivity
is noticeably larger than the perpendicular diffusivity, contrary to the
expectations for an isotropic system. Moreover, we can extract a factor between
the two of approximately $1.5$ over the whole range, even in the limit $\gs
\approx 50$, \ie $\frac{1}{2} \gamma_0^2 \approx 32$. This suggests that the
bare horizontal diffusivity $D_{x,0}$ grows with shear rate. We confirmed that
this behavior is not affected by boundary conditions by  repeating the
simulations in a quadratic box where we found the same increase.

\begin{figure}
	\center
	\includegraphics[width=\linewidth]{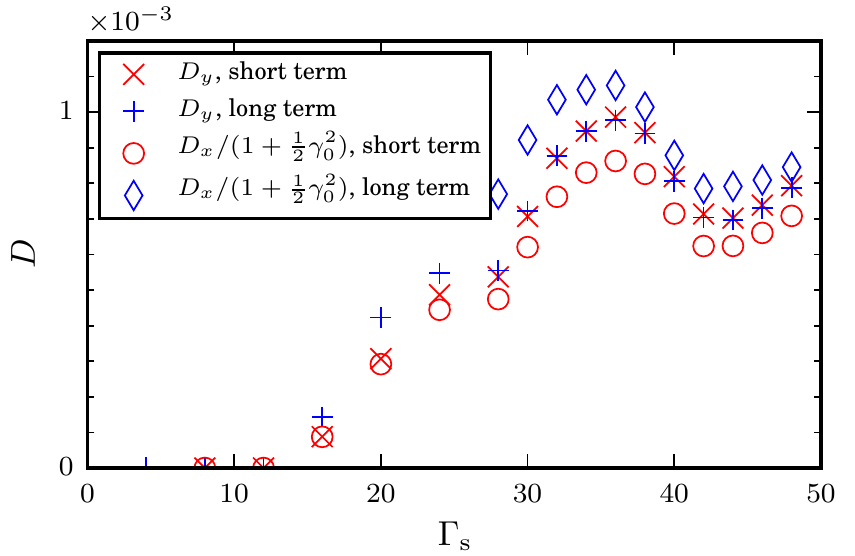}
	\caption[Diffusion constants, collapsed, $\gi = 1.0$]{
		(Color online)
		(Rescaled) diffusivities at $\gi = 1.0$. Below $\gs = 28$, the
		simulation has not converged towards the asymptotic ordered state: While most
		particles are arranged in a lattice structure, multiple defects exist which
		slowly grow out. Therefore, the long term diffusivity strongly deviates from
		the short term diffusivity and is neglected. At larger shear rates, we observe
		good agreement with \eqnref{eq:chSheareqCollapse}, \ie the anisotropy is
		caused by advection-diffusion coupling.
		\label{fig:chShearfigCollapse10}}
\end{figure}

The differences between bare longitudinal and transverse diffusion disappear for
stronger interactions $\gi$. The diffusivities for $\gi=1.0$ in an
ensemble of 900 particles are presented in \figref{fig:chShearfigCollapse10}. On
the numerical side, we note that due to the strong interaction and the resulting
steeper gradients, integration times increase considerably. Below $\gs =28$
the system has not yet converged towards an ordered state: While most particles
are arranged in a lattice configuration, unordered regions exist which relax
only very slowly. This is also reflected in the squared displacements: Instead
of approaching a linear growth for larger $\gamma t$, they obey a power law,
$\langle \Delta y^2 \rangle = 2D_y (\gamma t)^\alpha $ with $\alpha > 1$. For
larger shear rates, the simulation has converged and both the short term and
long term diffusivity perpendicular to the strain coincide. Virtually the same
applies to the longitudinal diffusivity, though we do not observe perfect
agreement. The cause for the deviation is the same as for $\gs = 0.1$, the
particles being correlated for a few periods before showing diffusive behavior.
The more important observation is that the rescaled diffusivities $D_{x,0}$ in
\figref{fig:chShearfigCollapse10} nicely bracket the perpendicular diffusivity.
Thus, for larger $\gi$ the anisotropy can be completely explained by
advection-diffusion coupling.

\subsection{Exploration of the $\gi$-$\gs$-parameter plane}
\begin{figure}
	\includegraphics[width=\linewidth]{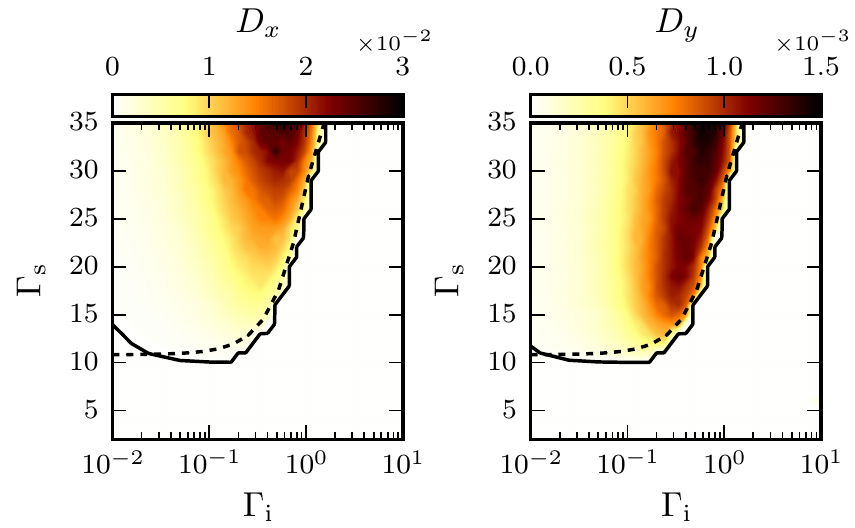}
	\caption[Diffusion constants]{
	  (Color online)
	  The long term diffusion constants $D_x$ (left panel) and $D_y$ (right panel)
	  for different values of $\gi$ and $\gs$. The diffusivity is obtained by
	  fitting data with \eqnref{eq:ch2eq2}. The black solid lines indicate $D_x =
	  1\e{-5}$ and $D_y=1\e{-6}$, respectively. The dashed lines are the smoothed
	  curves taken from \figref{fig:Ordering}, indicating the loss of structural
	  ordering. Results for $\gs > 5 \cdot 2\pi$ have to be treated with care due
	  to the limited box-width of $L_x = 10$: At this point, during one period each
	  particle slips beyond five particles in each direction. This leads to a
	  stabilizing effect for $\gi > 0.5$: Although the shear force is strong, at
	  the turning points of the oscillation the particle interaction still
	  dominates and promotes ordering into a lattice configuration. This behavior
	  is not observed for larger box-sizes.
  	\label{fig:fig1ch1}
 	}
\end{figure}

Next, we turn to the dependence on the shear rate and the interaction strength.
In \figref{fig:fig1ch1}, both parallel and perpendicular diffusivities are shown
in a large domain of the parameter space $\gi \times \gs$.

We observe a distinct boundary where diffusivities become finite as the shear
rate $\gs$ is increased. It nicely corresponds to the boundary determined by the
loss of ordering in the system, observed in \figref{fig:Ordering}. For a better
comparability, a smoothed representation of the curve where the ordering number
$|\Psi_k|$ drops below $0.4$ is shown in \figref{fig:fig1ch1} as well.

In general, the diffusivity increases with the interaction parameter $\gi$ up to
a critical value where it drops to zero. A small diffusivity $D$ for small
values of the interaction parameter $\gi$ is plausible, since a relatively
higher shear rate $\gs$ is needed to bring particles closer together and to
allow the system to display large spatial fluctuations. The dominant force is
the external forcing, and after one period the particles are very close to where
they were one period before. This leads to a small diffusion constant. In fact,
the relaxation time becomes so long that we cannot measure a proper diffusion
constant. However, there are qualitative differences depending on the shear rate
$\gs$: For small shear rates, the particles are evenly distributed, but
unordered, whereas for larger shear rates the particles are uniformly
distributed. This is also supported by two-particle correlation functions.
However, the transition between the two regimes is not as sharp as for stronger
particle interactions. For large interaction parameters we see a similar
behavior, but for a different reason. Since the interaction is comparatively
strong, we need a high shear rate to keep the particles away from their
equilibrium positions. On the other hand, for large $\gi$, the settling time
scale is shorter, and since the period of the external forcing is fixed to $1$,
we expect to find a critical region where particles settle into an ordered state
before they can start moving in a random fashion. This is reflected in a
relatively sharper transition in \figref{fig:fig1ch1} for $\gi\approx 1$. As
would be anticipated, stronger shear rates are required as the particle
interaction is increased.

For a dilute solution of particles, \citet{Pine2005} found that the diffusivity
is independent of the period of the forcing. In order to check whether our
results are independent of the driving period, we have to vary the period while
keeping the strain $\gamma$ fixed. With our dimensionless parameters, this
corresponds to a variation of $\gi$, while $\gs$ is kept fixed. From
\figref{fig:fig1ch1}, we see that for small $\gi \leq 0.2$, \ie short enough
periods and weak interactions,  the onset of a finite diffusivity and the loss
of ordering is independent of the driving period. For longer periods and
stronger forces though, the interaction between particles has more time to drive
the system to the ordered, reversible state.

However, the relation to the experiments of \citet{Pine2005} is more
complicated. In the experiments, the shear rate was fixed and the shear
amplitude was changed by varying the period. This implies a change in $\gi$ as
well, because the particles have more time to interact. In the fluids context,
the interaction comes from the hydrodynamic forces, and they become smaller if
the shear rate (and thus the speed) is reduced. The parameter regime where both
systems show similar phenomenology then is the range of small $\gi$, where the
transition is determined by $\gs$ alone.


\section{Linear stability analysis}\label{sec:Floquet}
\begin{figure}
  \includegraphics[width=\linewidth]{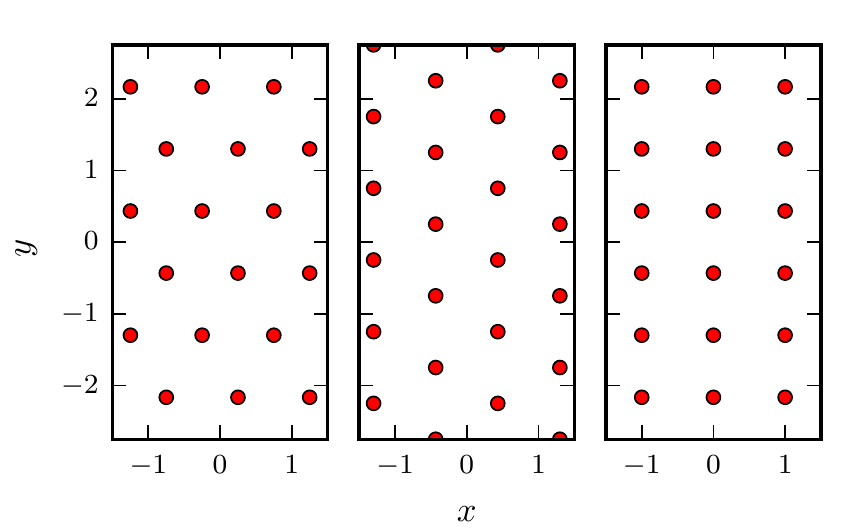}
  \caption{\label{fig:ch3fig1}
  	(Color online)
  	The three particle configurations considered in the linear stability
  	analysis, from left to right: equilateral triangular lattice aligned along
  	the $x$- and $y$-axis and rectangular lattice. These are also the asymptotic
  	states found in the dynamical system.
  }
\end{figure}

Investigation of the long term trajectories of simulations in the reversible
regime show that they approach regular lattices. That regular lattices are force
equilibria, even under periodic shear, follows from the fact that to each
particle pair there is a corresponding one with opposite forces. More formally,
consider a lattice configuration with a point symmetry in particle spacings,
i.e. a situation where for every pair $i$, $j$ one can find another pair $i$,
$k$, such that $\x_{ij} = -\x_{ik} $. Then it is easy to see that the
inter-particle forces in \eqnref{eq:ch1eq5} vanish. Since the shear force only
introduces an affine transformation, the force balance stays unaffected during
the shear process. Therefore, the regular lattices under shear correspond to a
periodic cycle of system \eqref{eq:ch1eq5},
\begin{equation}
\label{eq:ch3eq1}
\x_i^o(t) = \x_i^{\mathrm{lattice}} + 
 \frac{\gs}{2\pi}y_i^{\mathrm{lattice}}\sin(2\pi t)\vek{e}_x,
\end{equation}
for all parameter values $\gi$ and $\gs$. The initial regular lattice
points are denoted $\x_i^{\mathrm{lattice}}$, and their time-dependent
cousins are $\x_i^o(t)$. The regular lattices we consider are the ones obtained
under small shear (\figref{fig:ch3fig1}): the equilateral triangular lattice
with particles aligned along the shear direction, the same lattice rotated by
$90^\circ$, and the rectangular lattice aligned with the shear. In order to
determine their stability against small perturbations, we have to perform a
Floquet analysis, on account of the periodic variation of the sheared lattices.

\subsection{Floquet analysis}
We linearize the evolution equation (\ref{eq:ch1eq5}) around the periodic cycle
(\ref{eq:ch3eq1})
\begin{equation}\label{eq:ch3eq2}
  \x_i(t) = \x_i^o(t) + \gv{\delta}_i(t).
\end{equation}
This yields a system of linear ordinary differential equations for the
perturbation vector $\gv{\delta}(t)$,
\begin{equation}\label{eq:ch3eq3}
 \deri{\gv{\delta}}{t} =  
  \gi\left( \mathcal{D}(t) - \mathcal{M}(t) \right)\gv{\delta} 
  + \gs\mathcal{S}(t)\gv{\delta} = \mathcal{W}(t)\gv{\delta}.
\end{equation}
The right-hand side of this system consists of a time-periodic coefficient
matrix $\mathcal{W}(t)$ composed of $2N \times 2N$ matrices
\begin{equation}\label{eq:ch3eq4}
\mathcal{D}(t) = 
\begin{pmatrix}
G_1 &		& 0	\\
    & \ddots 	&	\\
 0  &		& G_n
\end{pmatrix},
\end{equation}
and
\begin{equation}\label{eq:ch3eq5}
\mathcal{M}(t) = 
\begin{pmatrix}
0 & B_{1,2} & \dotsb & B_{1,N}\\
B_{1,2} & 0 & \dotsb & B_{2,N}\\
\vdots & \vdots & \ddots & \vdots\\
B_{1,N} & B_{2,N} & \dotsb & 0
\end{pmatrix},
\end{equation}
where $G_i(t) = \sum_{j\neq i}B_{i,j}(t)$ with the $2 \times 2$ Jacobi matrices
$B_{i,j}(t)$ of the pairwise particle interactions,
\begin{equation}\label{eq:ch3eq6}
B_{i,j}(t) =  \frac{\openone_2}{{\|\vek x_i^o - \vek x_j^o\|}^4} 
 - \frac{4  \left(\vek x_i^o - \vek x_j^o\right)\otimes 
 	\left(\vek x_i^o - \vek x_j^o \right)}{{\|\vek x_i^o - \vek x_j^o\|}^6}.
\end{equation}
Here, $\otimes$ denotes the dyadic product and $\openone_2$ is the
identity-matrix in two dimensions. Finally, we have the Jacobi matrix of the
external forcing,
\begin{equation}\label{eq:ch3eq7}
  \mathcal{S}(t) = 
   \begin{pmatrix}
    S_2(t) &	 & 0\\
        & \ddots & \\
    0   &	 & S_2(t)
   \end{pmatrix} , \quad 
  S_2(t) = 
  \begin{pmatrix}
    0 & \cos(2\pi t) \\
    0 & 0
  \end{pmatrix}.
\end{equation}

Since the coefficient matrix has a periodic time-dependence, the linear
stability analysis requires an integration of the equations over a full period
\cite{Chicone}. To this end, we compute the principal fundamental matrix
$\Phi(t)$ of system (\ref{eq:ch3eq3}) with initial condition
$\Phi(0)=\openone_{2N}$. The matrix after a full period, $\Phi(1) \equiv \Phi$,
is the Floquet matrix and is not necessarily a symmetric matrix. The stability
properties of system (\ref{eq:ch3eq3}) then depend on the eigenvalues of $\Phi$
only. The eigenvalues $\sigma$ of $\Phi$ are called Floquet multipliers. For
stability, we require $\lvert \sigma \rvert$  for all possible eigenvalues
$\sigma$ of $\Phi$ to be less than one. There are two neutral modes ($\sigma =
1$), connected with the invariance of perturbations along the orbit and
translational invariance in the $x$-direction.

As an important technical detail we mention that the boundary conditions have to
be implemented such that they are compatible with the symmetries of the system so that the
symmetries of the matrices are preserved. To this end, we introduce an
interaction radius around each particle, and take all interactions with
particles inside the circle into account. The radius is chosen to be at least of
the order of the box dimensions.

\subsection{Stable and unstable configurations}

\begin{figure}
	\includegraphics[width=\linewidth]{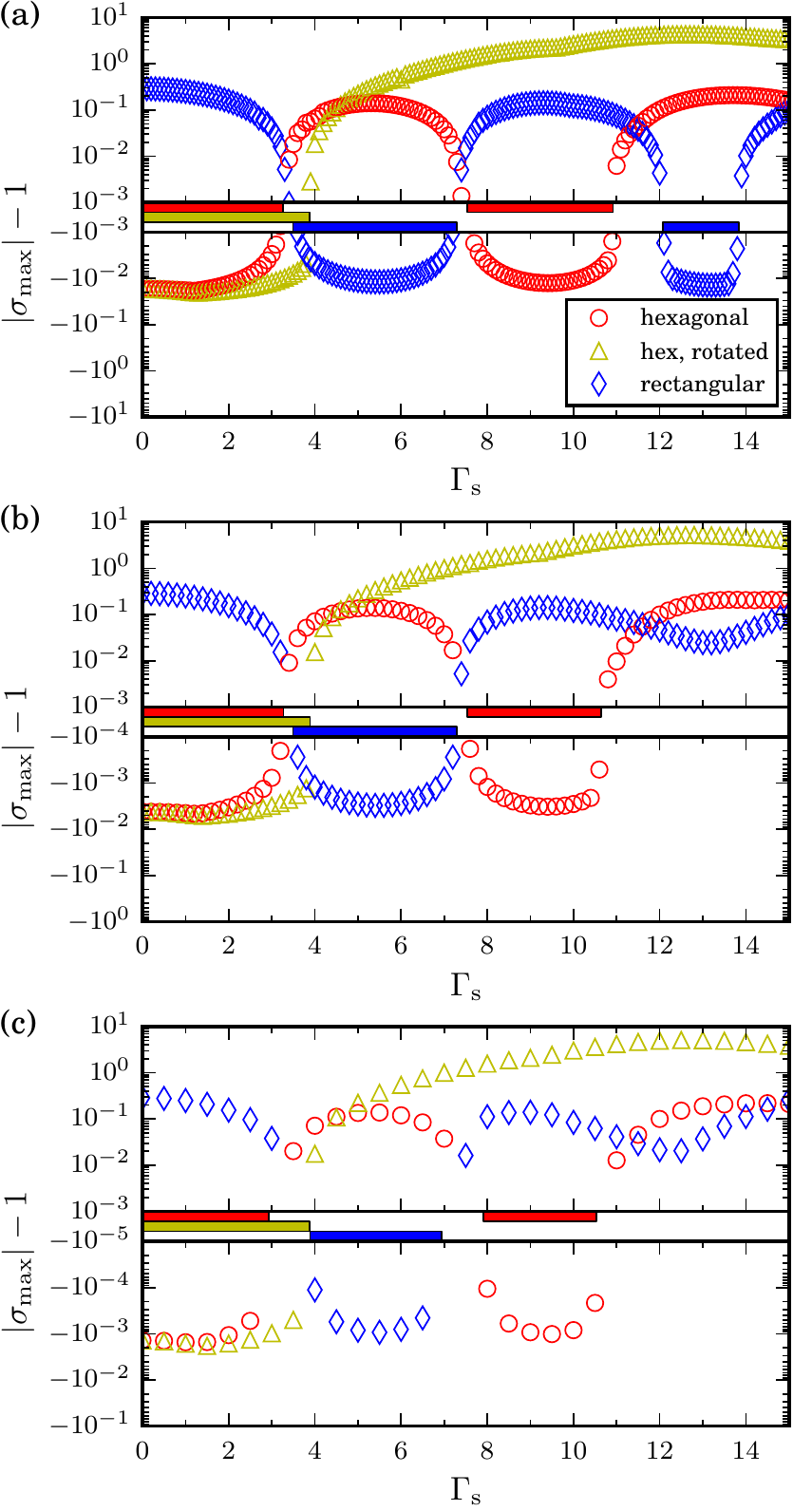}
	\caption{
		(Color online)
		The largest eigenvalues $\sigma_{\max}$ of $\Phi$ in dependence on $\gs$
		for the three investigated configurations with $N=100$, $400$, and $900$
		particles from top to bottom.  Neutral modes are omitted. In each upper panel,
		markers indicate eigenvalues $|\sigma_{\max}|>1$ and thus unstable
		configurations, and stable modes ($|\sigma_{\max}|<1$) in the lower panels
		accordingly. In between, the regions of stability are represented by colored
		bars.  We find that for values of $\gs \lesssim 11$, linearly stable
		states exist. Beyond, linear stability is lost. This is in good agreement with
		the diffusion constant calculated earlier. With increased number of particles,
		the stable modes become less stable, the qualitative difference between $400$
		and $900$ particles being rather marginal. The stable patch at $\gs>12$
		in the 100-particle system can clearly be identified as a finite size effect
		due to the limited box-width. The value of $\gi$ is set to $0.1$ in all
		calculations.
		\label{fig:figFloquet}
	}
\end{figure}

We studied the stability of configurations with $N = 100,400$ and $900$
particles. We find that results are qualitatively the same, with the stable
modes being closer to unity for more particles. We investigated three different
configurations as described above, and for the rotated hexagonal configuration
we swapped $L_x$ and $L_y$.

It turns out that for $\gi = 0.1$, at least one configuration is stable for
values of $\gs \lesssim 10.5$. This value corresponds to a shear rate where each
particle is shifted by $1.5$ particle distances in relation to the neighboring
rows. For larger values, the periodic cycles remain unstable. The largest
multipliers are shown in \figref{fig:figFloquet} for a wide range of the shear
rate. We find three regions with different stable configurations. For $\gs
\lesssim 3.5$, both the hexagonal state and its rotated counterpart are stable.
Thereafter, the rectangular state is stable up to a shear rate of $\gs \lesssim
7.5$ until finally the hexagonal configuration is stable again. The changes
occur at shear rates corresponding to relative displacements of neighboring rows
of approximately $0.5$, $1$ and $1.5$ particle distances. They nicely correspond
to the modulations of the orientation number shown in \figref{fig:Ordering}.
Only in a small interval at $\gs\approx 7.5$, all considered lattices appear to
be unstable. The inspection of long term trajectories shows, that the system
relaxes towards a mixture of both the hexagonal and rectangular lattice and
still is in the reversible regime.

The different behavior of the rotated hexagonal configuration can be understood
from a geometrical point of view: In the hexagonal lattice with particles
aligned along the shear direction, the initial  configuration is reproduced
after a shear displacement by one particle distance. In the rotated lattice a
shift by four particle distances is needed before the lattice returns,
corresponding to a shear rate of $\gs \approx 20$. An analysis of the
Jacobian indicates that it picks up strongly unstable modes during this process.

We do not find any qualitative differences
between the $400$ and $900$ particle ensemble. Quantitatively, the stable modes
become less stable with increasing number of particles. Additionally, the points
where the configurations change stability are slightly shifted. The magnitudes
of the unstable modes, however, are almost identical. The 100 particle system,
on the other hand, clearly shows signs of boundary effects. In the range $12
\lesssim \gs \lesssim 14$, the rectangular lattice becomes stable
again. This is neither observed in the larger systems, nor in the simulations.
The source might be the limited box width in conjunction with the long range
potential. Except for  this detail, the 100 particle system qualitatively agrees
with the larger ensembles as well.

A variation of the interaction strength $\gi$ reveals that the location of
stable regions is independent thereof for $\gi \lesssim 0.2$. For a larger
interaction strength, more stable regions emerge, in general accordance with
\figref{fig:fig1ch1}. However, the computational demand considerably increases
as $\gi$ is increased, so that a systematic investigation with an adequately
large number of particles is unfeasible.


\section{Particle distributions}\label{sec:Dist}
\subsection{Radial distribution functions}
\begin{figure}
	\center
	\includegraphics[width=\linewidth]{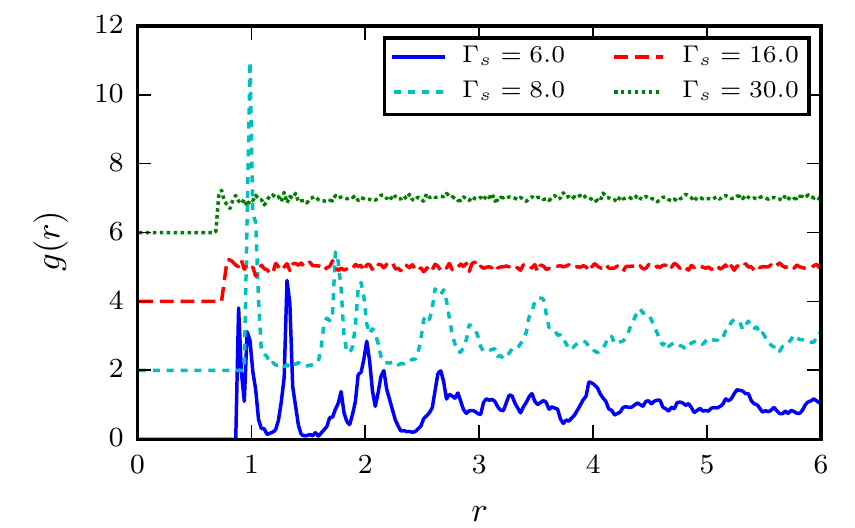}
	\caption[Radial correlation function]{
		(Color online)
		Radial correlation functions of ensembles at different shear rates. For
		clearer distinction, graphs are shifted by a constant. At low shear rates
		($\gs = 6.0$ and $8.0$, respectively), the crystalline structure is
		reflected by isolated peaks at the next neighbor distances: At $\gs=6.0$
		the rectangular lattice with $a=1.0$ and $b=\sqrt{3}/2$, and the hexagonal
		lattice with $a=1.0$ at $\gs = 8.0$. At higher shear rates, in the
		irreversible regime, the correlation function shows features of a liquid:
		Beyond a minimum distance, the function quickly approaches $1.0$ and displays
		only small fluctuations. Simulation of 900 particles, $\gi = 0.1$.
	\label{fig:chShearfigCorrelation}}
\end{figure}

Additional information about the structure of the system that goes beyond the
value of the diffusion constants and reversibility can be obtained through
correlation functions that contain information about the arrangements of
particles. The radially averaged two-particle correlation function counts the
number of particles within a ring bounded by the radii $r-\delta r$ and
$r+\delta r$, normalized by the mean number of particles in such a ring,
\begin{equation}
	g(r) = \frac{1}{N}
	\frac{\int_{r-\delta r/2}^{r+\delta r/2}\sum_{ij}\delta(r_{ij}-\tilde{r})
		 d \tilde{r}}
	{\int_{r-\delta r/2}^{r+\delta r/2} 2\pi \tilde{r}\rho d \tilde{r}},
\end{equation}
where $\rho=N/(L_xL_y)$ is the average particle density of the system. For a
crystalline structure, one expects sharp peaks corresponding to the underlying
lattice, the first one at the next neighbor distance, and so on. In a liquid
state no long range ordering is found so that the correlation approaches one at
larger distances. As neighboring particles repel each other, one expects a gap
close to the particle and a strong peak at the position of the nearest
neighbors. \figref{fig:chShearfigCorrelation} shows the correlation functions of
the system for different shear rates $\gs$ (and $\Gamma=0.1$). To obtain
better statistics all simulations in this section are done for systems of 900
particles. Inspection of the final states and the displacements per particle 
show that they behave exactly as the smaller systems of 100 particles. Each
ensemble is taken at a full period after a total simulation time of 11000
periods, i.e. we assume the system has reached its asymptotic state, which is
certainly the case for the systems at larger shear rates.

As expected, at low shear rates the ordered structure of the state is reflected
in the correlation function.  At $\gs = 6.0$, we find two superimposed
peaks at $r=\sqrt{3}/2$ and $1.0$, corresponding to the rectangular lattice. At
a slightly increased shear rate of $\gs = 8.0$ the asymptotic state
changes, as can be seen in the correlation function which exhibits isolated
peaks at $r=1.0$ and $\sqrt{3}$, corresponding to the next neighbor distances in
a hexagonal lattice. At higher shear rates after the motion became irreversible,
the correlation function shows features of a liquid: Beyond a minimum distance
owing to the repelling potential, the function quickly approaches 1.0 and shows
only small modulations. For even larger shear rates, the correlation function
becomes featureless except for the next neighbor repulsion: the system is in a
gas-like state. However, no spatial information can be extracted in the latter
cases. The next neighbor distance is identified as $0.8$ at $\gs = 16.0$,
and it decreases to $0.7$ at $\gs = 30.0$. This points to an improvement of
the mixing process with increasing shear rate, which leaves the particles less
time to relax and to return to larger separations.

\subsection{2-d correlation functions}

\begin{figure}
	\center
	\includegraphics[width=\columnwidth]{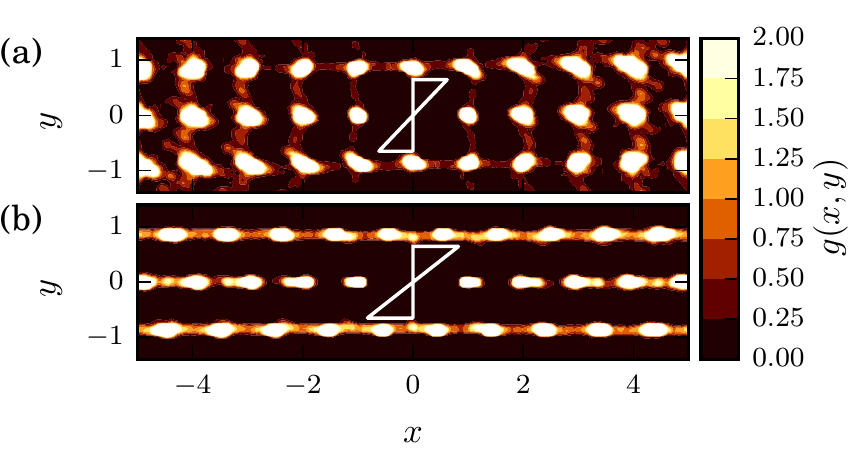}
	\caption[2d correlation function, pt. I]{
		(Color online)
		The 2-dimensional correlation function $g(x,y)$ at low shear rates. Colors
		indicate the neighbor density. At the white patches, $g(x,y)$ exceeds $2.0$.
		Just as the radial correlation function $g(r)$, $g(x,y)$ reflects the
		crystalline structure: At $\gs = 6.0$, (a), we observe the rectangular
		lattice structure, whereas at $\gs = 8.0$, (b), the hexagonal lattice is
		recovered. In both cases, the blur is caused by minor lattice defects whereby
		in the latter case almost perfect vertical alignment is observed. The white
		triangles indicate how far the populated rows have been sheared against each
		other over the past quarter period. They are set at arbitrary height such that
		features of the correlation function are not occluded. Simulation of 900
		particles, $\gi = 0.1$
		\label{fig:chShearfigCorrelation2d_1}}
\end{figure}

In order to obtain more spatial information, we now turn to the spatially
resolved two-particle correlation function
\begin{equation}
g(x,y) = \frac{1}{n} 
\iint_{\substack{x-\delta x/2\\y-\delta y/2}}^
 {\substack{y+\delta y/2\\x+\delta x/2}}
 \sum_{ij}\delta(x_{ij}-\tilde{x})\delta(y_{ij}-\tilde{y}) 
 d \tilde{x} d \tilde{y},
\end{equation}
where $n=N\rho \delta x \delta y$. Just as its radial counterpart, the 2d
correlation function relates the number of particles in a rectangular box of
size $\delta x \times \delta y$ around the distance $x$ and $y$ to the mean
number of particles expected in such a region.

Figure \ref{fig:chShearfigCorrelation2d_1} shows the 2d correlations for the
ordered configurations at low shear rates. Data was acquired in the same manner
as for the radial correlation function. In the plot, the correlation function is
color coded from dark brown to white with increasing density and centered around
the neutral density of unity. As expected from the radial correlation, for
$\gs = 6.0$ we find a rectangular lattice with most particles scattered
closely to the lattice points and a few particles found on the grid lines. This
is due to some minor lattice defects. Accordingly, at $\Gamma = 8.0$ we recover
the hexagonal lattice. Again, particles gather closely to the lattice sites,
at which vertical alignment is almost perfect. The horizontal scatter is due to
over- or under-population in rows which locally lead to rectangular arrangements
of particles. Note that both ensembles show mirror symmetries along the $x$- and
$y$-axis, i.e., there is no memory in the system if it has been sheared to the
left or to the right over the last half period.

\begin{figure}
	\center
	\includegraphics[width=\columnwidth]{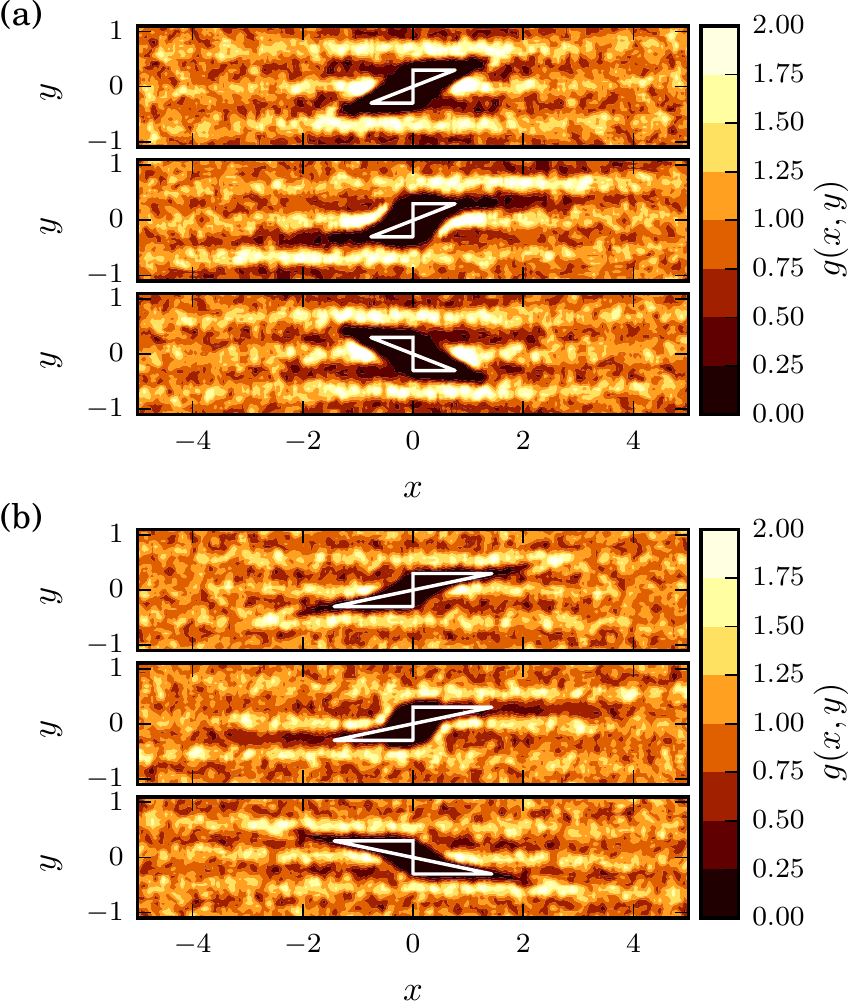}
	\caption[2d correlation function, pt. II]{
		(Color online)
		The 2-dimensional correlation function $g(x,y)$ at high shear rates over half
		a period at (a)  $\gs  =16.0$ and (b) $\gs = 30.0$. Colors indicate
		the particle density. At the white patches, $g(x,y)$ exceeds $2.0$. Each
		sub-figure shows,  from top to bottom, snapshots at $t=0$, $t=T/4$, and
		$T=T/2$. We observe a break-down of the mirror symmetry found at lower shear
		rates. While the free space is tilted to the right at the beginning of a
		period, half a period later it is tilted to the opposite direction. Moreover,
		we find remnants of the hexagonal lattice in the form of  stripes of increased
		density spaced by roughly the layer distance of the crystal. The white
		triangles indicate how far the populated rows have been sheared against each
		other over the past quarter period. They are set at arbitrary height such that
		features of the correlation function are not occluded. Simulation of 900
		particles, $\gi = 0.1$
	\label{fig:chShearfigCorrelation2d_2}}
\end{figure}

\begin{figure*}
	\includegraphics{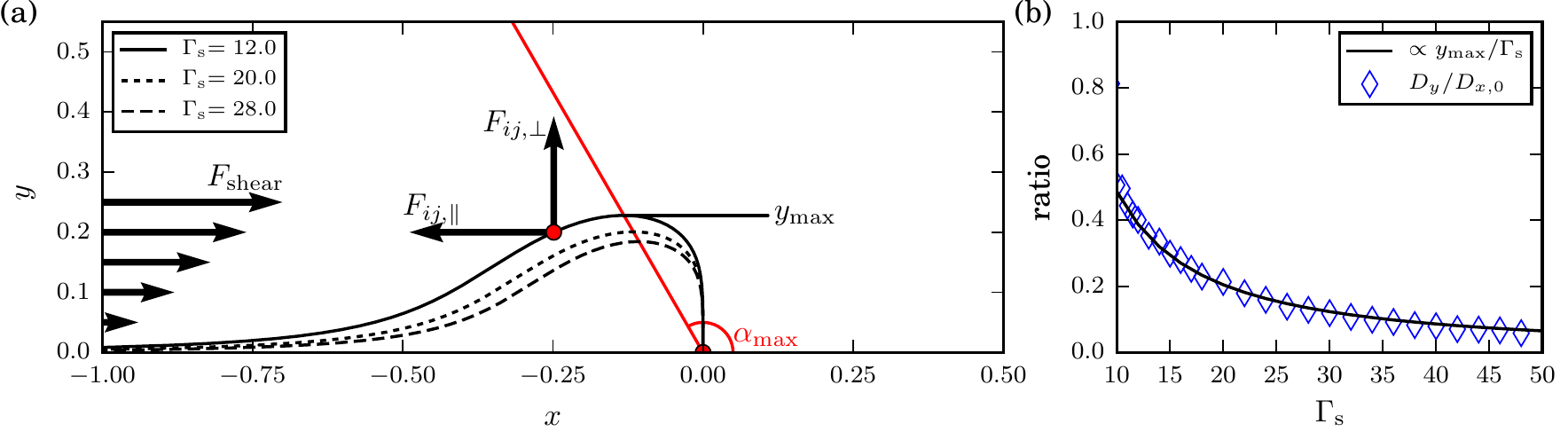}
	\caption{
			Formation of a shear rate dependent depletion zone (a) and ratio of the bare
			diffusivities and length scales in dependence on the shear rate (b). In (a),
			we consider two particles (\rlap{\color{red}$\bullet$}$\circ$), one being
			fixed at the origin, and the other one being subject to the forces (arrows)
			due to particle interaction $\vek F_{ij}$ at $\gi=0.1$ and the shear $\vek
			F_{\mathrm{shear}}$, which linearly increases with the separation in the
			$y$-direction. Black lines represent the points where horizontal forces are
			balanced. Below, the free particle moves to the left, \ie particles outside
			the area cannot enter it. With increasing shear rate, the maximum value
			$y_{\max}$ decreases. The red line marks the position of largest thickness of
			the depletion zone, changing with $\gs$. In (b), the ratio of the bare
			diffusivities $D_y/D_{x,0}$ shows the same behavior as the quotient
			$y_{\max{}}/\gs$.
		\label{fig:Banding}}
\end{figure*}

The picture changes upon advancing to higher shear rates. In
\figref{fig:chShearfigCorrelation2d_2} are shown the 2d correlations $g(x,y)$
across half a period at $\gs = 16.0$ and $30.0$, respectively. In figure (a) are
shown snapshots of $\gs = 16.0$  at the beginning of a period, at the turning
point of $t=T/4$ when the flow has stopped and strain is maximal, and after half
a period at $t=T/2$, when the flow has reversed its direction and strain is zero
again. The most prominent feature is the break-down of the symmetry. At the
start of the period, $t=0$, we observe a dark brown rhombic region around the
origin where no other particles are found. It is oriented to the right. As would
be expected from the symmetry of the flow, half a period later we observe a
similar rhombus, but this time oriented to the left. In between, at the turning
point of the flow, its outer boundary at $y \approx \pm 0.5$ is stretched along
the strain even more, while at the centerline $y=0$ it keeps its boundaries. In
effect, this leads to less populated protrusions above and below the centerline,
concurrent with the relative strain. They closely resemble the gaps found in
models describing the emergence of moonlet propellers in Saturn's A ring
\cite{Spahn2000,*Sremcevic2002,*Sremcevic2007}. A \textquoteleft
propeller-shaped\textquoteright \ region was also described by \citet{Keim2013c}
in their study of the model in \cite{Corte2008}. Due to a different definition
of the shear protocol, our snapshot at $t=T/4$ corresponds to $t=T/2$ in their
system, and by symmetry relates to $t=T$. A remarkable difference to their
observation is that they describe the correlations for the reversible state,
whereas we here find it in the irreversible state. Therefore, also the irregular
state maintains a phase memory of the shear process, at least for the
interactions studied here.

We can understand this phenomenon when starting at $t=-T/4$. At this time, the
flow is at its left turning point. We now look at the line $x=0$. At short
distances, particles on the left of this line are repelled to the left, and over
the next quarter period, the flow is not fast  (i.e. strong) enough to pull them
beyond $x=0$. Thus, at $t=0$ we observe the rhombus: it simply corresponds to
the line $x=0$, advected over the last quarter period. This suppression of
transport continues up to $t=T/4$. Since this does not apply to the centerline,
the rhombic shape gets distorted and we observe the aforementioned protrusions.
As soon as the low density area leaves the core area, particles from the other
sides are pushed into the void. This is reflected by a slight increase in
$g(x,y)$ towards the tips of the protrusions. However, at $t=T/2$, the remnants
are still visible (brown (gray) areas in the upper right and lower left).

The maxima of the correlation function are found at $x \approx \pm1.0$ and $y=0$
at every time step. This corresponds to a strong localization of next neighbors
parallel to the shear. Additionally, in the perpendicular direction we find a
modulation in particle density, with maxima at $y \approx 0$ and $\pm
0.6\dots0.8$, and multiples thereof. The modulations extent over several
particle distances parallel to the shear before they become weaker and decay.
This has two implications. First, although the system is far from its
equilibrium of a hexagonal lattice configuration, it still shows the vertical
density modulation corresponding to the layers in the lattice. With increasing
shear rate, this becomes less pronounced, but is still detectable. Second, those
maxima in the correlation function hint at a general anisotropy: The next
neighbor distance parallel to the shear is conspicuously larger than the next
neighbor distance in the perpendicular direction, as reflected in the location
of the layers. 

At the higher shear rate $\gs=30.0$, shown in
\figref{fig:chShearfigCorrelation2d_2}(b), the picture qualitatively stays the
same. We still observe the density modulations and the flow dependent variation
of the next neighbor distances. What changes is the shape of the particle-free
space. At the beginning of the period, it gets stretched even more, but does not
resemble a rhombus any more. This is due to the fixed neighbor relations along
the centerline at $x \approx \pm 0.8\dots1.2$.  At the turning point, $t=T/4$,
the free space is a slightly tilted oval instead of a rounded rhombus. As
expected, the density modulations in the perpendicular direction have a smaller
wavelength than at $\gs=16.0$, and the first maxima are found at $y \approx
\pm 0.5\dots0.6$.

The reduction of the vertical spacing with shear rate can be explained within a
simple model. To this end, we consider two particles, a fixed one placed at the
origin and a free one, as sketched in  \figref{fig:Banding}(a). For now, we
replace the periodic shear by a constant one which linearly increases with the
separation $y$. Thus, the free particle is subject to two different forces, the
inter-particle force $\vek F_{ij}$ and the shear force $\vek F_{\mathrm{shear}}$
acting along the $x$-direction. Taking only forces parallel to the shear into
account, we can compute the particle position where horizontal forces balance,
\begin{align}
	F_\mathrm{shear} &= - F_{ij,\parallel}.
\end{align}
If the mobile particle  is close to the $x$-axis, forces are balanced at large
distances. For larger separations in $y$, the shear force dominates and the
force from the stationary particle is too weak to balance them. The curves of
horizontal force equilibrium are shown in \figref{fig:Banding}(a) for various
shear rates. Particles in the area below the curves move to the left and are
deflected upwards until they are finally far enough away to pass the stationary
particle. This deflection can also be observed in the correlation function in
\figref{fig:chShearfigCorrelation2d_2} at $t=T/4$, where the maximum at $x=\pm1$
is bent away from the centerline. Due to the periodic repetition and the
many-particle interactions, this eventually leads to the formation of a somewhat
larger depletion zone where $g(x,y)$ is very low. When increasing the shear
rate, forces are balanced closer to the origin and so the area of the depletion
zone decreases. The maximum separation $y_{\max}$ varies along a straight line
with $\alpha_{\max} = 2\pi/3$ the angle towards the $x$-axis, and approaches the
origin in the limit of very large shear rates. By contrast, the hard-core
interactions used by \citet{Keim2013c} do not lead to a shear rate dependent
separation: They report a constant separation in the perpendicular direction
with boundaries at $y=\pm1$.

Advection-diffusion coupling \eqref{eq:chSheareqCollapse} is based upon the
assumption that a single particle performs diffusive motion. However, in the
system studied in this paper we have a multi-particle interaction and thus the
effects of the surrounding particles have to be taken into account. These
interactions determine the anisotropies of the diffusivities in
\figref{fig:chShearfigCollapse}. To see this, note that the free path of a
particle is determined by the regions of low particle density, \ie low $g(x,y)$.
From the previous observations and the simple model, we can derive two
assumptions: First, the free path perpendicular to the shear is governed by the
maximum separation $y_{\max}$. Obviously, they do not correspond directly to
each other since $y_{\max}$ is smaller than the separation in
\figref{fig:chShearfigCorrelation2d_2}. Yet, $y_{\max}$ should determine the
general dependency on the shear rate. Second, the free path parallel to the
shear is set by the low-density protrusions, which roughly increase along with
the shear rate $\gs$.

These assumptions then suggest that the ratio of the bare diffusivities
$D_y/D_{x,0}$ should be proportional to $y_{\max{}}/\gs$.
\figref{fig:Banding}(b) shows that this is indeed the case for $\gs > 15$. This
shows that the unexpected anisotropy of the system is connected with the
emergence of two different length scales which change with the shear rate.

\section{Summary}\label{sec:sum}

We have seen that even a model as simple as (\ref{eq:ch1eq5}) displays features
usually associated with much more complex systems. We found chaotic and ordered
behavior which can occur in different regions of the parameter space
$(\gi, \gs)$: The transition is similar for each value of the
interaction strength: For small shear strains, after some transient time the
system becomes reversible, the shear can be considered as a small perturbation.
Beyond an interaction-dependent critical shear rate, this feature is lost and we
observe chaotic motion. For stronger interactions, the critical point is delayed
to higher shear rates.

We identified the diffusivity as an indicator of chaotic motion. As would be
expected for a diffusive system subject to shear, we observe advection-diffusion
coupling: The parallel diffusivity is strongly enhanced as compared to
perpendicular diffusivity. For large interaction strengths, we can explain this
anisotropy using \eqnref{eq:chSheareqCollapse}. For low interactions though we
observe additional anisotropic effects, which we can attribute to the
2d-correlations of particles: Whereas the free space parallel to the shear
increases with the shear rate, the next neighbor distance perpendicular to the
shear decreases with increasing shear rate. Both effects combined allow us to
explain the relative anisotropy of the bare diffusivities. Additionally, we were
able to identify a phase dependency of the particle distributions, which is
concealed when looking upon \eg stroboscopic maps.

Finally, we could relate the critical shear rate associated with the loss of
reversibility to the stability properties of sheared regular lattices. So at
least for the present system the onset of irreversibility is connected with an
instability of the asymptotic, regular state.

\bibliography{bibliography}

\begin{thebibliography}{46}%
\makeatletter
\providecommand \@ifxundefined [1]{%
 \@ifx{#1\undefined}
}%
\providecommand \@ifnum [1]{%
 \ifnum #1\expandafter \@firstoftwo
 \else \expandafter \@secondoftwo
 \fi
}%
\providecommand \@ifx [1]{%
 \ifx #1\expandafter \@firstoftwo
 \else \expandafter \@secondoftwo
 \fi
}%
\providecommand \natexlab [1]{#1}%
\providecommand \enquote  [1]{``#1''}%
\providecommand \bibnamefont  [1]{#1}%
\providecommand \bibfnamefont [1]{#1}%
\providecommand \citenamefont [1]{#1}%
\providecommand \href@noop [0]{\@secondoftwo}%
\providecommand \href [0]{\begingroup \@sanitize@url \@href}%
\providecommand \@href[1]{\@@startlink{#1}\@@href}%
\providecommand \@@href[1]{\endgroup#1\@@endlink}%
\providecommand \@sanitize@url [0]{\catcode `\\12\catcode `\$12\catcode
  `\&12\catcode `\#12\catcode `\^12\catcode `\_12\catcode `\%12\relax}%
\providecommand \@@startlink[1]{}%
\providecommand \@@endlink[0]{}%
\providecommand \url  [0]{\begingroup\@sanitize@url \@url }%
\providecommand \@url [1]{\endgroup\@href {#1}{\urlprefix }}%
\providecommand \urlprefix  [0]{URL }%
\providecommand \Eprint [0]{\href }%
\providecommand \doibase [0]{http://dx.doi.org/}%
\providecommand \selectlanguage [0]{\@gobble}%
\providecommand \bibinfo  [0]{\@secondoftwo}%
\providecommand \bibfield  [0]{\@secondoftwo}%
\providecommand \translation [1]{[#1]}%
\providecommand \BibitemOpen [0]{}%
\providecommand \bibitemStop [0]{}%
\providecommand \bibitemNoStop [0]{.\EOS\space}%
\providecommand \EOS [0]{\spacefactor3000\relax}%
\providecommand \BibitemShut  [1]{\csname bibitem#1\endcsname}%
\let\auto@bib@innerbib\@empty
\bibitem [{\citenamefont {Hahn}(1950)}]{Hahn1950}%
  \BibitemOpen
  \bibfield  {author} {\bibinfo {author} {\bibfnamefont {E.~L.}\ \bibnamefont
  {Hahn}},\ }\href {\doibase 10.1103/PhysRev.80.580} {\bibfield  {journal}
  {\bibinfo  {journal} {Phys. Rev.}\ }\textbf {\bibinfo {volume} {80}},\
  \bibinfo {pages} {580} (\bibinfo {year} {1950})}\BibitemShut {NoStop}%
\bibitem [{\citenamefont {Carr}\ and\ \citenamefont
  {Purcell}(1954)}]{Carr1954}%
  \BibitemOpen
  \bibfield  {author} {\bibinfo {author} {\bibfnamefont {H.~Y.}\ \bibnamefont
  {Carr}}\ and\ \bibinfo {author} {\bibfnamefont {E.~M.}\ \bibnamefont
  {Purcell}},\ }\href {\doibase 10.1103/PhysRev.94.630} {\bibfield  {journal}
  {\bibinfo  {journal} {Phys. Rev.}\ }\textbf {\bibinfo {volume} {94}},\
  \bibinfo {pages} {630} (\bibinfo {year} {1954})}\BibitemShut {NoStop}%
\bibitem [{\citenamefont {Homsy}(2008)}]{Homsy2008}%
  \BibitemOpen
  \bibinfo {editor} {\bibfnamefont {G.~M.}\ \bibnamefont {Homsy}},\ ed.,\
  \href@noop {} {\emph {\bibinfo {title} {{Multimedia Fluid Dynamics}}}},\
  \bibinfo {edition} {2nd}\ ed.\ (\bibinfo  {publisher} {Cambridge University
  Press},\ \bibinfo {year} {2008})\BibitemShut {NoStop}%
\bibitem [{\citenamefont {Niggemeier}\ \emph {et~al.}(1993)\citenamefont
  {Niggemeier}, \citenamefont {von Plessen}, \citenamefont {Sauter},\ and\
  \citenamefont {Thomas}}]{Niggemeier1993}%
  \BibitemOpen
  \bibfield  {author} {\bibinfo {author} {\bibfnamefont {W.}~\bibnamefont
  {Niggemeier}}, \bibinfo {author} {\bibfnamefont {G.}~\bibnamefont {von
  Plessen}}, \bibinfo {author} {\bibfnamefont {S.}~\bibnamefont {Sauter}}, \
  and\ \bibinfo {author} {\bibfnamefont {P.}~\bibnamefont {Thomas}},\ }\href
  {\doibase 10.1103/PhysRevLett.71.770} {\bibfield  {journal} {\bibinfo
  {journal} {Phys. Rev. Lett.}\ }\textbf {\bibinfo {volume} {71}},\ \bibinfo
  {pages} {770} (\bibinfo {year} {1993})}\BibitemShut {NoStop}%
\bibitem [{\citenamefont {Chaiken}\ \emph {et~al.}(1986)\citenamefont
  {Chaiken}, \citenamefont {Chevray}, \citenamefont {Tabor},\ and\
  \citenamefont {Tan}}]{Chaiken1986}%
  \BibitemOpen
  \bibfield  {author} {\bibinfo {author} {\bibfnamefont {J.}~\bibnamefont
  {Chaiken}}, \bibinfo {author} {\bibfnamefont {R.}~\bibnamefont {Chevray}},
  \bibinfo {author} {\bibfnamefont {M.}~\bibnamefont {Tabor}}, \ and\ \bibinfo
  {author} {\bibfnamefont {Q.~M.}\ \bibnamefont {Tan}},\ }\href {\doibase
  10.1098/rspa.1986.0115} {\bibfield  {journal} {\bibinfo  {journal} {Proc. R.
  Soc. A}\ }\textbf {\bibinfo {volume} {408}},\ \bibinfo {pages} {165}
  (\bibinfo {year} {1986})}\BibitemShut {NoStop}%
\bibitem [{\citenamefont {Roberts}\ and\ \citenamefont
  {Quispel}(1992)}]{Roberts1992}%
  \BibitemOpen
  \bibfield  {author} {\bibinfo {author} {\bibfnamefont {J.~A.~G.}\
  \bibnamefont {Roberts}}\ and\ \bibinfo {author} {\bibfnamefont {G.~R.~W.}\
  \bibnamefont {Quispel}},\ }\href {\doibase 10.1016/0370-1573(92)90163-T}
  {\bibfield  {journal} {\bibinfo  {journal} {Phys. Rep.}\ }\textbf {\bibinfo
  {volume} {216}},\ \bibinfo {pages} {63} (\bibinfo {year} {1992})}\BibitemShut
  {NoStop}%
\bibitem [{\citenamefont {Casati}\ \emph {et~al.}(1986)\citenamefont {Casati},
  \citenamefont {Chirikov}, \citenamefont {Guarneri},\ and\ \citenamefont
  {Shepelyansky}}]{Casati1986}%
  \BibitemOpen
  \bibfield  {author} {\bibinfo {author} {\bibfnamefont {G.}~\bibnamefont
  {Casati}}, \bibinfo {author} {\bibfnamefont {B.~V.}\ \bibnamefont
  {Chirikov}}, \bibinfo {author} {\bibfnamefont {I.}~\bibnamefont {Guarneri}},
  \ and\ \bibinfo {author} {\bibfnamefont {D.~L.}\ \bibnamefont
  {Shepelyansky}},\ }\href {\doibase 10.1103/PhysRevLett.56.2437} {\bibfield
  {journal} {\bibinfo  {journal} {Phys. Rev. Lett.}\ }\textbf {\bibinfo
  {volume} {56}},\ \bibinfo {pages} {2437} (\bibinfo {year}
  {1986})}\BibitemShut {NoStop}%
\bibitem [{\citenamefont {Pastawski}\ \emph {et~al.}(1995)\citenamefont
  {Pastawski}, \citenamefont {Levstein},\ and\ \citenamefont
  {Usaj}}]{Pastawski1995}%
  \BibitemOpen
  \bibfield  {author} {\bibinfo {author} {\bibfnamefont {H.~M.}\ \bibnamefont
  {Pastawski}}, \bibinfo {author} {\bibfnamefont {P.~R.}\ \bibnamefont
  {Levstein}}, \ and\ \bibinfo {author} {\bibfnamefont {G.}~\bibnamefont
  {Usaj}},\ }\href {\doibase 10.1103/PhysRevLett.75.4310} {\bibfield  {journal}
  {\bibinfo  {journal} {Phys. Rev. Lett.}\ }\textbf {\bibinfo {volume} {75}},\
  \bibinfo {pages} {4310} (\bibinfo {year} {1995})}\BibitemShut {NoStop}%
\bibitem [{\citenamefont {Eckhardt}(2003)}]{Eckhardt2003}%
  \BibitemOpen
  \bibfield  {author} {\bibinfo {author} {\bibfnamefont {B.}~\bibnamefont
  {Eckhardt}},\ }\href {\doibase 10.1088/0305-4470/36/2/306} {\bibfield
  {journal} {\bibinfo  {journal} {J. Phys. A}\ }\textbf {\bibinfo {volume}
  {36}},\ \bibinfo {pages} {371} (\bibinfo {year} {2003})}\BibitemShut
  {NoStop}%
\bibitem [{\citenamefont {Pine}\ \emph {et~al.}(2005)\citenamefont {Pine},
  \citenamefont {Gollub}, \citenamefont {Brady},\ and\ \citenamefont
  {Leshansky}}]{Pine2005}%
  \BibitemOpen
  \bibfield  {author} {\bibinfo {author} {\bibfnamefont {D.~J.}\ \bibnamefont
  {Pine}}, \bibinfo {author} {\bibfnamefont {J.~P.}\ \bibnamefont {Gollub}},
  \bibinfo {author} {\bibfnamefont {J.~F.}\ \bibnamefont {Brady}}, \ and\
  \bibinfo {author} {\bibfnamefont {A.~M.}\ \bibnamefont {Leshansky}},\ }\href
  {\doibase 10.1038/nature04380} {\bibfield  {journal} {\bibinfo  {journal}
  {Nature}\ }\textbf {\bibinfo {volume} {438}},\ \bibinfo {pages} {997}
  (\bibinfo {year} {2005})}\BibitemShut {NoStop}%
\bibitem [{\citenamefont {Guasto}\ \emph {et~al.}(2010)\citenamefont {Guasto},
  \citenamefont {Ross},\ and\ \citenamefont {Gollub}}]{Guasto2010}%
  \BibitemOpen
  \bibfield  {author} {\bibinfo {author} {\bibfnamefont {J.~S.}\ \bibnamefont
  {Guasto}}, \bibinfo {author} {\bibfnamefont {A.~S.}\ \bibnamefont {Ross}}, \
  and\ \bibinfo {author} {\bibfnamefont {J.~P.}\ \bibnamefont {Gollub}},\
  }\href {\doibase 10.1103/PhysRevE.81.061401} {\bibfield  {journal} {\bibinfo
  {journal} {Phys. Rev. E}\ }\textbf {\bibinfo {volume} {81}},\ \bibinfo
  {pages} {061401} (\bibinfo {year} {2010})}\BibitemShut {NoStop}%
\bibitem [{\citenamefont {Metzger}\ and\ \citenamefont
  {Butler}(2012)}]{Metzger2012}%
  \BibitemOpen
  \bibfield  {author} {\bibinfo {author} {\bibfnamefont {B.}~\bibnamefont
  {Metzger}}\ and\ \bibinfo {author} {\bibfnamefont {J.~E.}\ \bibnamefont
  {Butler}},\ }\href {\doibase 10.1063/1.3685537} {\bibfield  {journal}
  {\bibinfo  {journal} {Phys. Fluids}\ }\textbf {\bibinfo {volume} {24}},\
  \bibinfo {pages} {021703} (\bibinfo {year} {2012})}\BibitemShut {NoStop}%
\bibitem [{\citenamefont {Jeanneret}\ and\ \citenamefont
  {Bartolo}(2014)}]{Jeanneret2014}%
  \BibitemOpen
  \bibfield  {author} {\bibinfo {author} {\bibfnamefont {R.}~\bibnamefont
  {Jeanneret}}\ and\ \bibinfo {author} {\bibfnamefont {D.}~\bibnamefont
  {Bartolo}},\ }\href {\doibase 10.1038/ncomms4474} {\bibfield  {journal}
  {\bibinfo  {journal} {Nat. Commun.}\ }\textbf {\bibinfo {volume} {5}},\
  \bibinfo {pages} {3474} (\bibinfo {year} {2014})}\BibitemShut {NoStop}%
\bibitem [{\citenamefont {Franceschini}\ \emph {et~al.}(2011)\citenamefont
  {Franceschini}, \citenamefont {Filippidi}, \citenamefont {Guazzelli},\ and\
  \citenamefont {Pine}}]{Franceschini2011}%
  \BibitemOpen
  \bibfield  {author} {\bibinfo {author} {\bibfnamefont {A.}~\bibnamefont
  {Franceschini}}, \bibinfo {author} {\bibfnamefont {E.}~\bibnamefont
  {Filippidi}}, \bibinfo {author} {\bibfnamefont {E.}~\bibnamefont
  {Guazzelli}}, \ and\ \bibinfo {author} {\bibfnamefont {D.~J.}\ \bibnamefont
  {Pine}},\ }\href {\doibase 10.1103/PhysRevLett.107.250603} {\bibfield
  {journal} {\bibinfo  {journal} {Phys. Rev. Lett.}\ }\textbf {\bibinfo
  {volume} {107}},\ \bibinfo {pages} {250603} (\bibinfo {year}
  {2011})}\BibitemShut {NoStop}%
\bibitem [{\citenamefont {Franceschini}\ \emph {et~al.}(2014)\citenamefont
  {Franceschini}, \citenamefont {Filippidi}, \citenamefont {Guazzelli},\ and\
  \citenamefont {Pine}}]{Franceschini2014}%
  \BibitemOpen
  \bibfield  {author} {\bibinfo {author} {\bibfnamefont {A.}~\bibnamefont
  {Franceschini}}, \bibinfo {author} {\bibfnamefont {E.}~\bibnamefont
  {Filippidi}}, \bibinfo {author} {\bibfnamefont {E.}~\bibnamefont
  {Guazzelli}}, \ and\ \bibinfo {author} {\bibfnamefont {D.~J.}\ \bibnamefont
  {Pine}},\ }\href {\doibase 10.1039/c4sm00555d} {\bibfield  {journal}
  {\bibinfo  {journal} {Soft Matter}\ }\textbf {\bibinfo {volume} {10}},\
  \bibinfo {pages} {6722} (\bibinfo {year} {2014})}\BibitemShut {NoStop}%
\bibitem [{\citenamefont {Keim}\ and\ \citenamefont
  {Arratia}(2013)}]{Keim2013a}%
  \BibitemOpen
  \bibfield  {author} {\bibinfo {author} {\bibfnamefont {N.~C.}\ \bibnamefont
  {Keim}}\ and\ \bibinfo {author} {\bibfnamefont {P.~E.}\ \bibnamefont
  {Arratia}},\ }\href {\doibase 10.1039/c3sm51014j} {\bibfield  {journal}
  {\bibinfo  {journal} {Soft Matter}\ }\textbf {\bibinfo {volume} {9}},\
  \bibinfo {pages} {6222} (\bibinfo {year} {2013})}\BibitemShut {NoStop}%
\bibitem [{\citenamefont {Keim}\ and\ \citenamefont
  {Arratia}(2014)}]{Keim2013}%
  \BibitemOpen
  \bibfield  {author} {\bibinfo {author} {\bibfnamefont {N.~C.}\ \bibnamefont
  {Keim}}\ and\ \bibinfo {author} {\bibfnamefont {P.~E.}\ \bibnamefont
  {Arratia}},\ }\href {\doibase 10.1103/PhysRevLett.112.028302} {\bibfield
  {journal} {\bibinfo  {journal} {Phys. Rev. Lett.}\ }\textbf {\bibinfo
  {volume} {112}},\ \bibinfo {pages} {028302} (\bibinfo {year}
  {2014})}\BibitemShut {NoStop}%
\bibitem [{\citenamefont {Metzger}\ and\ \citenamefont
  {Butler}(2010)}]{Metzger2010}%
  \BibitemOpen
  \bibfield  {author} {\bibinfo {author} {\bibfnamefont {B.}~\bibnamefont
  {Metzger}}\ and\ \bibinfo {author} {\bibfnamefont {J.~E.}\ \bibnamefont
  {Butler}},\ }\href {\doibase 10.1103/PhysRevE.82.051406} {\bibfield
  {journal} {\bibinfo  {journal} {Phys. Rev. E}\ }\textbf {\bibinfo {volume}
  {82}},\ \bibinfo {pages} {051406} (\bibinfo {year} {2010})}\BibitemShut
  {NoStop}%
\bibitem [{\citenamefont {D\"{u}ring}\ \emph {et~al.}(2009)\citenamefont
  {D\"{u}ring}, \citenamefont {Bartolo},\ and\ \citenamefont
  {Kurchan}}]{During2009}%
  \BibitemOpen
  \bibfield  {author} {\bibinfo {author} {\bibfnamefont {G.}~\bibnamefont
  {D\"{u}ring}}, \bibinfo {author} {\bibfnamefont {D.}~\bibnamefont {Bartolo}},
  \ and\ \bibinfo {author} {\bibfnamefont {J.}~\bibnamefont {Kurchan}},\ }\href
  {\doibase 10.1103/PhysRevE.79.030101} {\bibfield  {journal} {\bibinfo
  {journal} {Phys. Rev. E}\ }\textbf {\bibinfo {volume} {79}},\ \bibinfo
  {pages} {030101} (\bibinfo {year} {2009})}\BibitemShut {NoStop}%
\bibitem [{\citenamefont {Cort\'{e}}\ \emph {et~al.}(2008)\citenamefont
  {Cort\'{e}}, \citenamefont {Chaikin}, \citenamefont {Gollub},\ and\
  \citenamefont {Pine}}]{Corte2008}%
  \BibitemOpen
  \bibfield  {author} {\bibinfo {author} {\bibfnamefont {L.}~\bibnamefont
  {Cort\'{e}}}, \bibinfo {author} {\bibfnamefont {P.~M.}\ \bibnamefont
  {Chaikin}}, \bibinfo {author} {\bibfnamefont {J.~P.}\ \bibnamefont {Gollub}},
  \ and\ \bibinfo {author} {\bibfnamefont {D.~J.}\ \bibnamefont {Pine}},\
  }\href {\doibase 10.1038/nphys891} {\bibfield  {journal} {\bibinfo  {journal}
  {Nat. Phys.}\ }\textbf {\bibinfo {volume} {4}},\ \bibinfo {pages} {420}
  (\bibinfo {year} {2008})}\BibitemShut {NoStop}%
\bibitem [{\citenamefont {Cort\'{e}}\ \emph {et~al.}(2009)\citenamefont
  {Cort\'{e}}, \citenamefont {Gerbode}, \citenamefont {Man},\ and\
  \citenamefont {Pine}}]{Corte2009}%
  \BibitemOpen
  \bibfield  {author} {\bibinfo {author} {\bibfnamefont {L.}~\bibnamefont
  {Cort\'{e}}}, \bibinfo {author} {\bibfnamefont {S.~J.}\ \bibnamefont
  {Gerbode}}, \bibinfo {author} {\bibfnamefont {W.}~\bibnamefont {Man}}, \ and\
  \bibinfo {author} {\bibfnamefont {D.~J.}\ \bibnamefont {Pine}},\ }\href
  {\doibase 10.1103/PhysRevLett.103.248301} {\bibfield  {journal} {\bibinfo
  {journal} {Phys. Rev. Lett.}\ }\textbf {\bibinfo {volume} {103}},\ \bibinfo
  {pages} {248301} (\bibinfo {year} {2009})}\BibitemShut {NoStop}%
\bibitem [{\citenamefont {Menon}\ and\ \citenamefont
  {Ramaswamy}(2009)}]{Menon2009}%
  \BibitemOpen
  \bibfield  {author} {\bibinfo {author} {\bibfnamefont {G.~I.}\ \bibnamefont
  {Menon}}\ and\ \bibinfo {author} {\bibfnamefont {S.}~\bibnamefont
  {Ramaswamy}},\ }\href {\doibase 10.1103/PhysRevE.79.061108} {\bibfield
  {journal} {\bibinfo  {journal} {Phys. Rev. E}\ }\textbf {\bibinfo {volume}
  {79}},\ \bibinfo {pages} {061108} (\bibinfo {year} {2009})}\BibitemShut
  {NoStop}%
\bibitem [{\citenamefont {Keim}\ and\ \citenamefont {Nagel}(2011)}]{Keim2011}%
  \BibitemOpen
  \bibfield  {author} {\bibinfo {author} {\bibfnamefont {N.~C.}\ \bibnamefont
  {Keim}}\ and\ \bibinfo {author} {\bibfnamefont {S.~R.}\ \bibnamefont
  {Nagel}},\ }\href {\doibase 10.1103/PhysRevLett.107.010603} {\bibfield
  {journal} {\bibinfo  {journal} {Phys. Rev. Lett.}\ }\textbf {\bibinfo
  {volume} {107}},\ \bibinfo {pages} {010603} (\bibinfo {year}
  {2011})}\BibitemShut {NoStop}%
\bibitem [{\citenamefont {Mohan}\ \emph {et~al.}(2013)\citenamefont {Mohan},
  \citenamefont {Pellet}, \citenamefont {Cloitre},\ and\ \citenamefont
  {Bonnecaze}}]{Mohan2013}%
  \BibitemOpen
  \bibfield  {author} {\bibinfo {author} {\bibfnamefont {L.}~\bibnamefont
  {Mohan}}, \bibinfo {author} {\bibfnamefont {C.}~\bibnamefont {Pellet}},
  \bibinfo {author} {\bibfnamefont {M.}~\bibnamefont {Cloitre}}, \ and\
  \bibinfo {author} {\bibfnamefont {R.}~\bibnamefont {Bonnecaze}},\ }\href
  {\doibase 10.1122/1.4802631} {\bibfield  {journal} {\bibinfo  {journal} {J.
  Rheol.}\ }\textbf {\bibinfo {volume} {57}},\ \bibinfo {pages} {1023}
  (\bibinfo {year} {2013})}\BibitemShut {NoStop}%
\bibitem [{\citenamefont {Regev}\ \emph {et~al.}(2013)\citenamefont {Regev},
  \citenamefont {Lookman},\ and\ \citenamefont {Reichhardt}}]{Regev2013}%
  \BibitemOpen
  \bibfield  {author} {\bibinfo {author} {\bibfnamefont {I.}~\bibnamefont
  {Regev}}, \bibinfo {author} {\bibfnamefont {T.}~\bibnamefont {Lookman}}, \
  and\ \bibinfo {author} {\bibfnamefont {C.}~\bibnamefont {Reichhardt}},\
  }\href {\doibase 10.1103/PhysRevE.88.062401} {\bibfield  {journal} {\bibinfo
  {journal} {Phys. Rev. E}\ }\textbf {\bibinfo {volume} {88}},\ \bibinfo
  {pages} {062401} (\bibinfo {year} {2013})}\BibitemShut {NoStop}%
\bibitem [{\citenamefont {Fiocco}\ \emph {et~al.}(2013)\citenamefont {Fiocco},
  \citenamefont {Foffi},\ and\ \citenamefont {Sastry}}]{Fiocco2013}%
  \BibitemOpen
  \bibfield  {author} {\bibinfo {author} {\bibfnamefont {D.}~\bibnamefont
  {Fiocco}}, \bibinfo {author} {\bibfnamefont {G.}~\bibnamefont {Foffi}}, \
  and\ \bibinfo {author} {\bibfnamefont {S.}~\bibnamefont {Sastry}},\ }\href
  {\doibase 10.1103/PhysRevE.88.020301} {\bibfield  {journal} {\bibinfo
  {journal} {Phys. Rev. E}\ }\textbf {\bibinfo {volume} {88}},\ \bibinfo
  {pages} {020301} (\bibinfo {year} {2013})}\BibitemShut {NoStop}%
\bibitem [{\citenamefont {Fiocco}\ \emph {et~al.}(2014)\citenamefont {Fiocco},
  \citenamefont {Foffi},\ and\ \citenamefont {Sastry}}]{Fiocco2014}%
  \BibitemOpen
  \bibfield  {author} {\bibinfo {author} {\bibfnamefont {D.}~\bibnamefont
  {Fiocco}}, \bibinfo {author} {\bibfnamefont {G.}~\bibnamefont {Foffi}}, \
  and\ \bibinfo {author} {\bibfnamefont {S.}~\bibnamefont {Sastry}},\ }\href
  {\doibase 10.1103/PhysRevLett.112.025702} {\bibfield  {journal} {\bibinfo
  {journal} {Phys. Rev. Lett.}\ }\textbf {\bibinfo {volume} {112}},\ \bibinfo
  {pages} {025702} (\bibinfo {year} {2014})}\BibitemShut {NoStop}%
\bibitem [{\citenamefont {Slotterback}\ \emph {et~al.}(2012)\citenamefont
  {Slotterback}, \citenamefont {Mailman}, \citenamefont {Ronaszegi},
  \citenamefont {van Hecke}, \citenamefont {Girvan},\ and\ \citenamefont
  {Losert}}]{Slotterback2012}%
  \BibitemOpen
  \bibfield  {author} {\bibinfo {author} {\bibfnamefont {S.}~\bibnamefont
  {Slotterback}}, \bibinfo {author} {\bibfnamefont {M.}~\bibnamefont
  {Mailman}}, \bibinfo {author} {\bibfnamefont {K.}~\bibnamefont {Ronaszegi}},
  \bibinfo {author} {\bibfnamefont {M.}~\bibnamefont {van Hecke}}, \bibinfo
  {author} {\bibfnamefont {M.}~\bibnamefont {Girvan}}, \ and\ \bibinfo {author}
  {\bibfnamefont {W.}~\bibnamefont {Losert}},\ }\href {\doibase
  10.1103/PhysRevE.85.021309} {\bibfield  {journal} {\bibinfo  {journal} {Phys.
  Rev. E}\ }\textbf {\bibinfo {volume} {85}},\ \bibinfo {pages} {021309}
  (\bibinfo {year} {2012})}\BibitemShut {NoStop}%
\bibitem [{\citenamefont {Schreck}\ \emph {et~al.}(2013)\citenamefont
  {Schreck}, \citenamefont {Hoy}, \citenamefont {Shattuck},\ and\ \citenamefont
  {O’Hern}}]{Schreck2013}%
  \BibitemOpen
  \bibfield  {author} {\bibinfo {author} {\bibfnamefont {C.~F.}\ \bibnamefont
  {Schreck}}, \bibinfo {author} {\bibfnamefont {R.~S.}\ \bibnamefont {Hoy}},
  \bibinfo {author} {\bibfnamefont {M.~D.}\ \bibnamefont {Shattuck}}, \ and\
  \bibinfo {author} {\bibfnamefont {C.~S.}\ \bibnamefont {O’Hern}},\ }\href
  {\doibase 10.1103/PhysRevE.88.052205} {\bibfield  {journal} {\bibinfo
  {journal} {Phys. Rev. E}\ }\textbf {\bibinfo {volume} {88}},\ \bibinfo
  {pages} {052205} (\bibinfo {year} {2013})}\BibitemShut {NoStop}%
\bibitem [{\citenamefont {Royer}\ and\ \citenamefont
  {Chaikin}(2014)}]{Royer2014}%
  \BibitemOpen
  \bibfield  {author} {\bibinfo {author} {\bibfnamefont {J.~R.}\ \bibnamefont
  {Royer}}\ and\ \bibinfo {author} {\bibfnamefont {P.~M.}\ \bibnamefont
  {Chaikin}},\ }\href {\doibase 10.1073/pnas.1413468112} {\bibfield  {journal}
  {\bibinfo  {journal} {Proc. Natl. Acad. Sci.}\ }\textbf {\bibinfo {volume}
  {112}},\ \bibinfo {pages} {49} (\bibinfo {year} {2014})}\BibitemShut
  {NoStop}%
\bibitem [{\citenamefont {Mangan}\ \emph {et~al.}(2008)\citenamefont {Mangan},
  \citenamefont {Reichhardt},\ and\ \citenamefont {{Olson
  Reichhardt}}}]{Mangan2008}%
  \BibitemOpen
  \bibfield  {author} {\bibinfo {author} {\bibfnamefont {N.}~\bibnamefont
  {Mangan}}, \bibinfo {author} {\bibfnamefont {C.}~\bibnamefont {Reichhardt}},
  \ and\ \bibinfo {author} {\bibfnamefont {C.~J.}\ \bibnamefont {{Olson
  Reichhardt}}},\ }\href {\doibase 10.1103/PhysRevLett.100.187002} {\bibfield
  {journal} {\bibinfo  {journal} {Phys. Rev. Lett.}\ }\textbf {\bibinfo
  {volume} {100}},\ \bibinfo {pages} {187002} (\bibinfo {year}
  {2008})}\BibitemShut {NoStop}%
\bibitem [{\citenamefont {Zhang}\ \emph {et~al.}(2010)\citenamefont {Zhang},
  \citenamefont {Zhou},\ and\ \citenamefont {Luo}}]{Zhang2010}%
  \BibitemOpen
  \bibfield  {author} {\bibinfo {author} {\bibfnamefont {W.}~\bibnamefont
  {Zhang}}, \bibinfo {author} {\bibfnamefont {W.}~\bibnamefont {Zhou}}, \ and\
  \bibinfo {author} {\bibfnamefont {M.}~\bibnamefont {Luo}},\ }\href {\doibase
  10.1016/j.physleta.2010.06.057} {\bibfield  {journal} {\bibinfo  {journal}
  {Phys. Lett. A}\ }\textbf {\bibinfo {volume} {374}},\ \bibinfo {pages} {3666}
  (\bibinfo {year} {2010})}\BibitemShut {NoStop}%
\bibitem [{\citenamefont {Okuma}\ \emph {et~al.}(2010)\citenamefont {Okuma},
  \citenamefont {Suzuki},\ and\ \citenamefont {Tsugawa}}]{Okuma2010}%
  \BibitemOpen
  \bibfield  {author} {\bibinfo {author} {\bibfnamefont {S.}~\bibnamefont
  {Okuma}}, \bibinfo {author} {\bibfnamefont {Y.}~\bibnamefont {Suzuki}}, \
  and\ \bibinfo {author} {\bibfnamefont {Y.}~\bibnamefont {Tsugawa}},\ }\href
  {\doibase 10.1016/j.physc.2009.10.067} {\bibfield  {journal} {\bibinfo
  {journal} {Phys. C}\ }\textbf {\bibinfo {volume} {470}},\ \bibinfo {pages}
  {S842} (\bibinfo {year} {2010})}\BibitemShut {NoStop}%
\bibitem [{\citenamefont {Motohashi}\ and\ \citenamefont
  {Okuma}(2011)}]{Motohashi2011}%
  \BibitemOpen
  \bibfield  {author} {\bibinfo {author} {\bibfnamefont {A.}~\bibnamefont
  {Motohashi}}\ and\ \bibinfo {author} {\bibfnamefont {S.}~\bibnamefont
  {Okuma}},\ }\href {\doibase 10.1088/1742-6596/302/1/012029} {\bibfield
  {journal} {\bibinfo  {journal} {J. Phys. Conf. Ser.}\ }\textbf {\bibinfo
  {volume} {302}},\ \bibinfo {pages} {012029} (\bibinfo {year}
  {2011})}\BibitemShut {NoStop}%
\bibitem [{\citenamefont {Okuma}\ \emph {et~al.}(2011)\citenamefont {Okuma},
  \citenamefont {Tsugawa},\ and\ \citenamefont {Motohashi}}]{Okuma2011}%
  \BibitemOpen
  \bibfield  {author} {\bibinfo {author} {\bibfnamefont {S.}~\bibnamefont
  {Okuma}}, \bibinfo {author} {\bibfnamefont {Y.}~\bibnamefont {Tsugawa}}, \
  and\ \bibinfo {author} {\bibfnamefont {A.}~\bibnamefont {Motohashi}},\ }\href
  {\doibase 10.1103/PhysRevB.83.012503} {\bibfield  {journal} {\bibinfo
  {journal} {Phys. Rev. B}\ }\textbf {\bibinfo {volume} {83}},\ \bibinfo
  {pages} {012503} (\bibinfo {year} {2011})}\BibitemShut {NoStop}%
\bibitem [{\citenamefont {Melzer}\ \emph
  {et~al.}(1996{\natexlab{a}})\citenamefont {Melzer}, \citenamefont {Homann},\
  and\ \citenamefont {Piel}}]{Melzer1996}%
  \BibitemOpen
  \bibfield  {author} {\bibinfo {author} {\bibfnamefont {A.}~\bibnamefont
  {Melzer}}, \bibinfo {author} {\bibfnamefont {A.}~\bibnamefont {Homann}}, \
  and\ \bibinfo {author} {\bibfnamefont {A.}~\bibnamefont {Piel}},\ }\href
  {\doibase 10.1103/PhysRevE.53.2757} {\bibfield  {journal} {\bibinfo
  {journal} {Phys. Rev. E}\ }\textbf {\bibinfo {volume} {53}},\ \bibinfo
  {pages} {2757} (\bibinfo {year} {1996}{\natexlab{a}})}\BibitemShut {NoStop}%
\bibitem [{\citenamefont {Melzer}\ \emph
  {et~al.}(1996{\natexlab{b}})\citenamefont {Melzer}, \citenamefont
  {Schweigert}, \citenamefont {Schweigert}, \citenamefont {Homann},
  \citenamefont {Peters},\ and\ \citenamefont {Piel}}]{Melzer1996a}%
  \BibitemOpen
  \bibfield  {author} {\bibinfo {author} {\bibfnamefont {A.}~\bibnamefont
  {Melzer}}, \bibinfo {author} {\bibfnamefont {V.~A.}\ \bibnamefont
  {Schweigert}}, \bibinfo {author} {\bibfnamefont {I.~V.}\ \bibnamefont
  {Schweigert}}, \bibinfo {author} {\bibfnamefont {A.}~\bibnamefont {Homann}},
  \bibinfo {author} {\bibfnamefont {S.}~\bibnamefont {Peters}}, \ and\ \bibinfo
  {author} {\bibfnamefont {A.}~\bibnamefont {Piel}},\ }\href {\doibase
  10.1103/PhysRevE.54.R46} {\bibfield  {journal} {\bibinfo  {journal} {Phys.
  Rev. E}\ }\textbf {\bibinfo {volume} {54}},\ \bibinfo {pages} {R46} (\bibinfo
  {year} {1996}{\natexlab{b}})}\BibitemShut {NoStop}%
\bibitem [{\citenamefont {Schweigert}\ \emph {et~al.}(1998)\citenamefont
  {Schweigert}, \citenamefont {Schweigert}, \citenamefont {Melzer},
  \citenamefont {Homann},\ and\ \citenamefont {Piel}}]{Schweigert1998}%
  \BibitemOpen
  \bibfield  {author} {\bibinfo {author} {\bibfnamefont {V.~A.}\ \bibnamefont
  {Schweigert}}, \bibinfo {author} {\bibfnamefont {I.~V.}\ \bibnamefont
  {Schweigert}}, \bibinfo {author} {\bibfnamefont {A.}~\bibnamefont {Melzer}},
  \bibinfo {author} {\bibfnamefont {A.}~\bibnamefont {Homann}}, \ and\ \bibinfo
  {author} {\bibfnamefont {A.}~\bibnamefont {Piel}},\ }\href {\doibase
  10.1103/PhysRevLett.80.5345} {\bibfield  {journal} {\bibinfo  {journal}
  {Phys. Rev. Lett.}\ }\textbf {\bibinfo {volume} {80}},\ \bibinfo {pages}
  {5345} (\bibinfo {year} {1998})}\BibitemShut {NoStop}%
\bibitem [{\citenamefont {Lees}\ and\ \citenamefont
  {Edwards}(1972)}]{Lees1972}%
  \BibitemOpen
  \bibfield  {author} {\bibinfo {author} {\bibfnamefont {A.~W.}\ \bibnamefont
  {Lees}}\ and\ \bibinfo {author} {\bibfnamefont {S.~F.}\ \bibnamefont
  {Edwards}},\ }\href {\doibase 10.1088/0022-3719/5/15/006} {\bibfield
  {journal} {\bibinfo  {journal} {J. Phys. C: Solid State Phys.}\ }\textbf
  {\bibinfo {volume} {5}},\ \bibinfo {pages} {1921} (\bibinfo {year}
  {1972})}\BibitemShut {NoStop}%
\bibitem [{Note1()}]{Note1}%
  \BibitemOpen
  \bibinfo {note} {See Supplemental Material at URL for movies.}\BibitemShut
  {Stop}%
\bibitem [{\citenamefont {Young}\ \emph {et~al.}(1982)\citenamefont {Young},
  \citenamefont {Rhines},\ and\ \citenamefont {Garrett}}]{Young1982}%
  \BibitemOpen
  \bibfield  {author} {\bibinfo {author} {\bibfnamefont {W.~R.}\ \bibnamefont
  {Young}}, \bibinfo {author} {\bibfnamefont {P.~B.}\ \bibnamefont {Rhines}}, \
  and\ \bibinfo {author} {\bibfnamefont {C.~J.~R.}\ \bibnamefont {Garrett}},\
  }\href {\doibase 10.1175/1520-0485(1982)012<0515:SFDIWA>2.0.CO;2} {\bibfield
  {journal} {\bibinfo  {journal} {J. Phys. Oceanogr.}\ }\textbf {\bibinfo
  {volume} {12}},\ \bibinfo {pages} {515} (\bibinfo {year} {1982})}\BibitemShut
  {NoStop}%
\bibitem [{\citenamefont {Chicone}(2006)}]{Chicone}%
  \BibitemOpen
  \bibfield  {author} {\bibinfo {author} {\bibfnamefont {C.}~\bibnamefont
  {Chicone}},\ }\href {\doibase 10.1007/0-387-35794-7} {\emph {\bibinfo {title}
  {{Ordinary Differential Equations with Applications}}}},\ \bibinfo {series}
  {Texts in Applied Mathematics}, Vol.~\bibinfo {volume} {34}\ (\bibinfo
  {publisher} {Springer New York},\ \bibinfo {year} {2006})\BibitemShut
  {NoStop}%
\bibitem [{\citenamefont {Spahn}\ and\ \citenamefont
  {Sremcevic}(2000)}]{Spahn2000}%
  \BibitemOpen
  \bibfield  {author} {\bibinfo {author} {\bibfnamefont {F.}~\bibnamefont
  {Spahn}}\ and\ \bibinfo {author} {\bibfnamefont {M.}~\bibnamefont
  {Sremcevic}},\ }\href@noop {} {\bibfield  {journal} {\bibinfo  {journal}
  {Astron. Astrophys.}\ }\textbf {\bibinfo {volume} {358}},\ \bibinfo {pages}
  {368} (\bibinfo {year} {2000})}\BibitemShut {NoStop}%
\bibitem [{\citenamefont {Sremcevic}\ \emph {et~al.}(2002)\citenamefont
  {Sremcevic}, \citenamefont {Spahn},\ and\ \citenamefont
  {Duschl}}]{Sremcevic2002}%
  \BibitemOpen
  \bibfield  {author} {\bibinfo {author} {\bibfnamefont {M.}~\bibnamefont
  {Sremcevic}}, \bibinfo {author} {\bibfnamefont {F.}~\bibnamefont {Spahn}}, \
  and\ \bibinfo {author} {\bibfnamefont {W.~J.}\ \bibnamefont {Duschl}},\
  }\href {\doibase 10.1046/j.1365-8711.2002.06011.x} {\bibfield  {journal}
  {\bibinfo  {journal} {Mon. Not. R. Astron. Soc.}\ }\textbf {\bibinfo {volume}
  {337}},\ \bibinfo {pages} {1139} (\bibinfo {year} {2002})}\BibitemShut
  {NoStop}%
\bibitem [{\citenamefont {Sremcevi\'{c}}\ \emph {et~al.}(2007)\citenamefont
  {Sremcevi\'{c}}, \citenamefont {Schmidt}, \citenamefont {Salo}, \citenamefont
  {Seiss}, \citenamefont {Spahn},\ and\ \citenamefont
  {Albers}}]{Sremcevic2007}%
  \BibitemOpen
  \bibfield  {author} {\bibinfo {author} {\bibfnamefont {M.}~\bibnamefont
  {Sremcevi\'{c}}}, \bibinfo {author} {\bibfnamefont {J.}~\bibnamefont
  {Schmidt}}, \bibinfo {author} {\bibfnamefont {H.}~\bibnamefont {Salo}},
  \bibinfo {author} {\bibfnamefont {M.}~\bibnamefont {Seiss}}, \bibinfo
  {author} {\bibfnamefont {F.}~\bibnamefont {Spahn}}, \ and\ \bibinfo {author}
  {\bibfnamefont {N.}~\bibnamefont {Albers}},\ }\href {\doibase
  10.1038/nature06224} {\bibfield  {journal} {\bibinfo  {journal} {Nature}\
  }\textbf {\bibinfo {volume} {449}},\ \bibinfo {pages} {1019} (\bibinfo {year}
  {2007})}\BibitemShut {NoStop}%
\bibitem [{\citenamefont {Keim}\ \emph {et~al.}(2013)\citenamefont {Keim},
  \citenamefont {Paulsen},\ and\ \citenamefont {Nagel}}]{Keim2013c}%
  \BibitemOpen
  \bibfield  {author} {\bibinfo {author} {\bibfnamefont {N.~C.}\ \bibnamefont
  {Keim}}, \bibinfo {author} {\bibfnamefont {J.~D.}\ \bibnamefont {Paulsen}}, \
  and\ \bibinfo {author} {\bibfnamefont {S.~R.}\ \bibnamefont {Nagel}},\ }\href
  {\doibase 10.1103/PhysRevE.88.032306} {\bibfield  {journal} {\bibinfo
  {journal} {Phys. Rev. E}\ }\textbf {\bibinfo {volume} {88}},\ \bibinfo
  {pages} {032306} (\bibinfo {year} {2013})}\BibitemShut {NoStop}%
\end{thebibliography}%

\typeout{===Height: \the\paperheight ===}%
\typeout{===Column width: \the\columnwidth ===}%
\typeout{===Line width: \the\linewidth ===}%
\typeout{===Text width: \the\textwidth ===}%

\end{document}